\documentclass[prb,twocolumn,showpacs]{revtex4}
\usepackage{graphicx}
\usepackage{amsmath}
\usepackage{amssymb}

\newcommand{\sss}{\scriptstyle}
\def\lsim{\
  \lower-1.2pt\vbox{\hbox{\rlap{$<$}\lower5pt\vbox{\hbox{$\sim$}}}}\ }
\def\gsim{\
  \lower-1.2pt\vbox{\hbox{\rlap{$>$}\lower5pt\vbox{\hbox{$\sim$}}}}\ }

\begin{document}

\title{To the theory of the electric activity of He II induced by waves of first and second sounds}

\author{Maksim D.~Tomchenko} \email{mtomchenko@bitp.kiev.ua}
\affiliation{Bogolyubov Institute for Theoretical Physics, 14-b Metrologicheskaya Street,
Kiev 03680, Ukraine}

\date{\today}
\begin{abstract}
An approximate microscopic model is proposed for the explanation of the electric signal
${\sss \triangle} U  \approx k_B {\sss \triangle} T/2e$ observed by A.\,S.~Rybalko in He II in the
experiments with standing half-wave of second sound. The model is based on the idea, due
to Gutlyanskii, of the one-directional polarization of $^4$He atoms located at the
electrode surface.  The calculated parameters of the electric signal are in approximate
agreement with the experimental ones. It is also predicted that a standing half-wave of
first sound should induce a variable signal with amplitude $ {\sss \triangle} U\sim {\sss \triangle}
p/(|e|n)\sim 3\cdot 10^{-5} {\sss \triangle} p\, \mbox{V} /\mbox{atm}$ at the electrode.
 It is shown also that the dependence of the polarizability of
   helium on temperature, $A(T)$, can be explained  if the tidal polarization
   of atoms is taken into account.  A possibility of the existence of the
   ``dry'' friction in He II at temperatures $T \lsim 0.5 \div 1 \,$K
      is discussed.
\end{abstract}

\maketitle

       \section{Introduction}

In the  experiment of  Rybalko \cite{rub}, a standing half-wave of second sound was generated  in He~II placed
in a metal resonator. In this case, a  variable potential difference between the
electrode on the resonator end (inside the resonator)  and the ground was registered.
This electric signal  oscillated with the frequency
 of the second sound and the amplitude ${\sss \triangle} U  \approx k_B {\sss \triangle}
 T/2e$, which was independent of the resonator sizes and the temperature, in the region $T=1.4\div 1.8\,$K. Here, $e$ is the electron charge,
  and ${\sss \triangle} T$
  is the amplitude of temperature oscillations in the
  second-sound wave.
This effect was rather unexpected since free $^4$He
atoms are neutral and do not possess electric dipole or higher multipole moments. After
six years since the first communication of the effect, it remains unexplained, in our opinion, in spite
of a number of attempts \cite{hod,kos,mel,push,lt1,volpol,gut,shev1,shev2,pash2010,min2010}.  The present
work is aimed at a microscopic explanation of the Rybalko effect.

He~II is electrically active apparently due to the tidal polarization of helium atoms
\cite{vol84,lt1} caused by interaction with neighbors \cite{wb1,wb2,lt2}. As shown in
Ref.~\onlinecite{volpol},
     the tidal mechanism causes bulk polarization of He~II
     by a wave of second sound. In this case, however,
     the electric signal is 1--2 orders of magnitude below the registered one and
     strongly depends on the temperature and resonator size, which is not observed
 in experiments. It is important that the signal should strongly depend on the resonator size
     for {\em any bulk mechanism\/} of ``spontaneous'' polarization of He~II \cite{lt1,volpol}.
     In the recent works \cite{pash2010,min2010}, some bulk models without any dependence of the resulting
signal on the resonator size were proposed.
But we do not agree with the way of the determination
   of the electric field there.
Below (see Appendix A), we discuss some methods
of calculation of the field for the given problem.

The idea that the experimental electric signal is related to the polarization of $^4$He
atoms located at the electrode surface was put forward in Ref.~\onlinecite{gut} and,
later on, in Refs.~\onlinecite{shev1,shev2}.  The surface atoms are polarized in the direction
perpendicular to the surface because of the asymmetry of forces acting on them in this
direction: helium atoms are on the one side, and atoms and electrons of metal are on the
other side.  However, only several layers of helium atoms are strongly polarized, and the
electric signal from them may be too weak (it was not calculated in
Refs.~\onlinecite{gut,shev1,shev2}). In Refs.~\onlinecite{shev1,shev2}, a general
analysis within the model of a gas consisting of small electron-hole pairs was performed.
However, such an approach is indirect and can be used at most only qualitatively because
a $^4$He atom possesses more complicated structure in comparison with an electron-hole
pair, and the mass of the atomic nucleus is much greater than the electron mass.

For purely geometric reasons, one can think of the following four sources of the signal:
\begin{enumerate}
\item the bulk polarization of He~II,

\item the near-surface polarization --- from regions in helium at distances from the
    electrode much larger than the interatomic distance but much smaller than the
    resonator size,

\item the surface polarization --- from several layers of $^4$He atoms at the
    electrode surface, and

\item the thermo-emf in the electrode.
\end{enumerate}

The bulk signal is excluded because the bulk polarization is modulated by the wave of
second sound. In this case, the larger is the wavelength $\lambda_{2}$ of second sound, the
greater is the distance at which the temperature difference is ``smeared''. Consequently,
the signal must strongly depend \cite{lt1} on $\lambda_{2}$ and, therefore, on the
resonator length $L_{r}$ as well, because of the relation $L_{r}=\lambda_{2}/2$.

The near-surface polarization hypothetically can be related to some motion of
quasiparticles at distances of hundreds or thousands of atomic layers from the electrode.
But we are unable to indicate a specific physical mechanism that could cause such a
motion under the conditions of the experiment under discussion.  Possibly, this source of
polarization can only be a part of the bulk polarization.

The thermo-emf contributes undoubtedly, but this contribution is apparently small.
Moreover, it must strongly depend on the electrode material for both  the contact
thermo-emf and the thermo-emf driven by the temperature gradient inside the electrode,
while the experiment showed that the signal is identical for three different electrodes.

Therefore, the polarization of helium at the electrode surface seems to be the most
probable effect.  It is this option that will be studied in detail in this paper. In
addition, we are going to calculate the parameters of the electric signal induced by a
standing half-wave of the {\em first\/} sound.

The main results of the work, except for Sec. III, V and Appendix B, were published in Ref.~\onlinecite{prb2011}.
Part of the results of Sec. II and IV was published also in Ref.~\onlinecite{dan2011}. The content of Sec. III was published in Ref.~\onlinecite{dryF}.

     \section{Polarization of a $^4$He atom located at the boundary between He~II and metal}

In this section, we calculate the dipole moment (DM) of a $^4$He atom located at a plane
boundary between He~II and metal. Let the metal and helium occupy the half-spaces $z<0$
and  $z>0$, respectively. The $z$ axis is directed into helium normally to the metal surface.
The DM of a single nonpolar atom at a distance $z_{0}$ from the plane surface of a metal
was calculated in Ref.~\onlinecite{m2}:
             \begin{equation}
  \textbf{d}_{\rm mir} = D_{4} |e| \frac{a_{B}^{5}}{z_{0}^4}
  \textbf{i}_{z},
  \label{3-1}     \end{equation}
     \begin{equation}
   D_{4} \approx \frac{3}{8} \frac{e^2}{\hbar
  a_{B}}\sum\limits_{n}\frac{\langle z_{n}^{4}+z_{n}^{2}x_{n}^{2}\rangle }{a_{B}^{4}}
  \frac{a}{b\omega_{a}(\omega_{a}+b)}.
  \label{3-2}     \end{equation}
     Here, the summation is performed over the electrons of an atom (for a $^4$He atom, $n=1,2$),
         $a=b^{2}=\omega_{\rm pl}^{2}/2$, $\omega_{\rm pl}$  is the plasma frequency of a metal,
         $\omega_{a}= {\sss \triangle} E/\hbar$ is the effective excitation frequency
   of the atom,  $x_{n}$ and $z_{n}$ are the
coordinates of the $n^{\textrm{th}}$ electron of the atom, $a_B =
\hbar^{2}/(me^2)=0.529\,\mbox{\AA}$, and the averaging is performed over the ground state
of the atom. Physically, the DM (\ref{3-1}) arises due to interaction of the atom with
its ``image'' in a metal  ``mirror,'' hence the notation $\textbf{d}_{\rm mir}$.  The
image is not an exact mirror image but represents a small perturbation of the
distribution of charges in a metal at large distances from the position of the exact
image.  However, this distribution is equivalent to an exact mirror image in its action,
which follows also from results in Ref.~\onlinecite{m3}. That work presents the study of the polarization of an atom
by a dielectric with the use of a more transparent method
  --- the DM was calculated as the sum of DMs induced by each of the
dielectric atoms with regard for the response; the result was identical to that
in Ref.~\onlinecite{m2}.

For a $^4$He atom, the quantity $D_{4}$ in Eq.~(\ref{3-2}) can be presented in the form
      \begin{equation}
  D_{4} \approx \frac{2Ry}{5 {\sss \triangle} E}\frac{\hbar \omega_{\rm pl}}{\hbar \omega_{\rm pl} + \sqrt{2} {\sss \triangle} E}
   \left \langle \frac{r^4}{a_{B}^4}\right \rangle,
  \label{3-3}     \end{equation}
   where $Ry=e^{2}/2a_{B}=13.6\,\mbox{eV}$,       and
    $ {\sss \triangle} E$ is the mean excitation energy
   of a $^4$He atom which is close  \cite{lt2} to the ionization energy \cite{gerc0} $ {\sss \triangle} E_{ion}
    \approx 24.58\,$eV.  For the simplest one-parameter
    wave function  $\Psi_{0}$ of the ground state of a $^4$He atom, we have $\langle r^4/a_{B}^4 \rangle \approx
    2.775$, whereas for the 80-parameter \cite{kin2} $\Psi_{0}$  close to the
    exact one, $\langle r^4/a_{B}^4 \rangle \approx
    3.973$ (see Ref.~\onlinecite{lt2}). The last value will be used below.

We now consider  a $^4$He atom located in the first layer of helium near the plane
surface of the  metal.  This atom has polarization $\textbf{d}_{\rm h}$, induced by all
other helium atoms.  It was shown in Refs.~\onlinecite{wb1,wb2} that two interacting
$^4$He atoms induce the following DM on each other:
     \begin{equation}
  \textbf{d} = -D_{7} |e| \frac{a_{B}^{8}}{R^7} \textbf{n},
  \label{3-6}     \end{equation}
        where $\textbf{n} =\textbf{R}/R$ is the unit vector along the direction to the
        neighboring  atom, and $D_{7}\approx 18.4$. In Ref.~\onlinecite{lt2},
     a similar formula with  $D_{7}\approx 25.2 \pm 2$ was obtained using a simpler method.
Taking into account both results, we assume
        \begin{equation}
  D_{7} \approx 23 \pm 5.
  \label{3-7}     \end{equation}

We note that equations (\ref{3-1}) and (\ref{3-6}) are obtained by neglecting the
exchange interaction and the higher corrections (with larger degrees of $z_{0}$  or $R$
in the denominator). Both approximations are valid at $z_{0}, R \gg a_{B}$. For He~II, we
have $z_{0}, R \gtrsim 3\,\mbox{\AA}$, which justifies the applicability of (\ref{3-1})
and (\ref{3-6}).

According to Eq.~(15) from Ref.~\onlinecite{wb1} and Eq.~(18) from Ref.~\onlinecite{lt2},
the quantity $\textbf{d}$, given by (\ref{3-6}), is proportional to the average of the DM
operator and to the \textit{square\/} of a perturbing potential (equal to the difference
of the total Hamiltonian of two interacting $^4$He atoms and the Hamiltonians of free
atoms, according to Eq.~(10) from  Ref.~\onlinecite{lt1}; this is the sum of the Coulomb
potentials). In this case, the exchange interaction is dropped in
Refs.~\onlinecite{wb1,lt2}, which is justified for the interatomic distances $\gtrsim
3\,\mbox{\AA}$ under consideration. Let us calculate  $\textbf{d}_{\rm h}$. The resulting
formula for $\textbf{d}_{\rm h}$ will include the square of the total perturbing
potential, equal to the sum of the perturbing potentials from each atom. The square of
the total potential is given by the sum of the squares of one-atom potentials (this gives
the sum of DMs induced by individual atoms) and the sum of cross terms, which also give a
nonzero contribution to the total DM $\textbf{d}_{\rm h}$.
 Therefore, the DM induced by the sum of atoms is not reduced to a sum of
 DMs induced by each atom separately.
We, however, neglect the corrections from cross terms,
which represent three-particle corrections and are usually neglected in many-particle problems.
Then the
polarization $\textbf{d}_{\rm h}$ from the collection of helium atoms is equal to the sum
of the polarizations $\textbf{d}_{{\rm h} j}$ (\ref{3-6}) from separate atoms:
           \begin{equation}
 \textbf{d}_{\rm h} \approx \sum\limits_{j}\textbf{d}_{{\rm h} j},
    \label{3-5}     \end{equation}
    and  the total DM of a $^4$He atom located at the metal surface  is equal to
           \begin{equation}
 \textbf{d}_{1} \approx \textbf{d}_{\rm h}+\textbf{d}_{\rm mir}.
    \label{3-4}     \end{equation}
The cross terms    renormalizes $\textbf{d}_{\rm h}$ at most by 20--30\%,  most likely,
        as compared with (\ref{3-5}).

        In view of (\ref{3-6}) and (\ref{3-7}), the sum  (\ref{3-5}) can be written
         in the form
         \begin{equation}
  \textbf{d}_{\rm h}  \approx   -d_{0}\int\limits_{0}^{2\pi}d\phi \int\limits_{0}^{\pi/2}d\theta \int\limits_{0}^{\infty}r^{2}dr
  \frac{g(\textbf{r})\bar{R}_{0}^{7}}{\bar{R}^{3}r^{7}}\frac{\textbf{r}}{r} \equiv -\textbf{i}_{z}\frac{d_{0}S_{7}}{4}\frac{n}{n_{0}},
         \label{3-8}     \end{equation}
    where
     \begin{equation}
  d_{0}   =   D_{7}|e|a_{B}^{8}/\bar{R}_{0}^{7} = 3.56\cdot 10^{-5}|e|a_{B},
         \label{3-9}     \end{equation}
     \begin{equation}
  S_{7}  =  \int\limits_{\Omega_{2}=4\pi} n_{0}g(\textbf{r})\frac{\bar{R}_{0}^{7}}{r^{7}}d\textbf{r},
            \label{3-10}     \end{equation}
     $n(T, p) = \bar{R}^{-3}(T, p)$ is the concentration of He~II,  $n_{0}=n(T=1\,K, \ p={\rm svp}) = \bar{R}_{0}^{-3}$,
      $\bar{R}_{0} = 3.578\,\mbox{\AA}$,
      and $g(\textbf{r})$ is the two-point correlation function,  i.e., the probability to find a helium atom
     at the displacement $\textbf{r}$ from another helium atom. If the density of helium is constant near the metal,
      then helium as the simple liquid is isotropic, and $g(\textbf{r})\equiv
     g(r)$.  In this case, $n\approx  n_{0}$ and $S_{7} = S_{7}(n_{0})\approx 14.9$ (see Ref.~\onlinecite{volpol}).
     At the saturation vapor pressure (svp), we obtain the polarization of the first layer of helium at the metal surface:
            \begin{equation}
  \textbf{d}_{\rm h}   \approx   -\textbf{i}_{z}\cdot 1.32 \cdot 10^{-4}|e|a_{B}.
         \label{3-11}     \end{equation}
         The polarization of the next layers is then determined by equation (\ref{new1}) which is derived below.

          \section{Properties of He II near the metal surface and a possibility of the ``dry'' friction}
      To calculate $\textbf{d}_{\rm h}$ (\ref{3-8}), (\ref{3-10}),
      it is necessary to know the state of helium at the metal surface, namely
      the density profile $\rho (T,p)$ and the function
      $g(\textbf{r})$. This state depends on the helium atom--
      metal potential, and  we must else consider the exhaustion of the superfluid (SF)
      component near the wall.

      Depending on the interaction potential between the metal and a
      $^4$He atom, helium near the metal surface can be in four
      different states \cite{cheng1,cheng2}. For the weakest potential (Cs),
      helium does not wet  the metal, and the first layer of helium near
      the metal is liquid.
      For stronger potentials (Rb, Na, K, Li), some wettability is
      present, but the first layer is still liquid.
            Most metals (Cu, Al, Au and others) have stronger interaction
      potential, making the first layer of helium solid, and
      the second one liquid.
           Finally, at the strongest possible
      potentials, two layers can become solid.
      The authors of Ref.~\onlinecite{cheng2} believe that this happens for Au.
      However, the potential for Au (see Ref.~\onlinecite{zar1})
      at the distance of two atomic layers is approximately
      equal to the interaction potential of two $^4$He atoms and cannot compress helium up to solidification. Therefore, we
      assume that only one layer of helium is solid for Au.

        To know the dependence $T(z)$ for helium at a wall, it is necessary to understand the properties of the
        SF-component near the wall.
        In the presence of the wetting, helium atoms adhere to the wall, but $\textbf{v}_{s}$ cannot continuously increase, as the
        distance from the wall increases. In this connection, V.L. Ginzburg  advanced the assumption \cite{ginz1}
        that $\textbf{v}_{s}$ has a discontinuity near the wall, and, therefore, the ``dry'' friction must be observed in helium-II. However,
        the experiment gave no evidence of such a friction \cite{gam}.
         This implies \cite{ginz2} that $\rho_{s}=0$ on the metal surface.
         To the best of our knowledge, the microscopic reason for such an
         exhaustion of $\rho_{s}$ is not clarified.
         Since $\rho_{s}$ is equal to $\rho - \rho_{n}$ by definition,
         the exhaustion of $\rho_{s}$ on the wall can be caused by the behavior of $\rho $ (i.e., that of atoms) or by the behavior of
         $\rho_{n}$ (i.e., that of quasiparticles). The first seems improbable --- $\rho $ ensures $\rho_{s}=0$ either
         due to $\rho =0$ (but the exact zero cannot be reached, since the wall does not represent the infinitely high energy barrier)
         or because of the exact equality $\rho = \rho_{n}$, whose validity is  improbable due to the properties of atoms ( $\rho$) ---
         there is no reason for atoms to be rearranged so that the relation $\rho = \rho_{n}$ be satisfied namely on the wall.
         It would be so that $\rho$ is very close to zero on the wall (i.e. $\rho_{s}$ is almost zero),
          but the dry friction arising in this case would be too low for the experimental registration. But,
          in this case,
           $\rho_{s}$ should be close to zero only at the wall at the distance which is significantly less than the mean interatomic one
          (indeed, nothing hampers atoms to approach one another so closely). However, the experiment \cite{3sound1,pat} shows that
          $\rho_{s}$ is close to zero at significantly larger distances from the wall equal approximately to two atomic
         layers. This is namely the effective radius of a roton \cite{rr}. Hence,  we may conclude that, most likely,
          $\rho_{s}=0$ due to the properties of quasiparticles.
         In other words, there is a certain reason for the concentration of quasiparticles to be maximal on the wall
         and for the condition of the $\lambda$-transition, $\rho_{s}=0$, to be realized.
                    Below, we will study this possibility.

           The following simple mechanism is possible. The microscopic calculation \cite{krot1,krot2}
           and the experiment \cite{2dr} imply that the energy $\Delta_{2D}$ of a surface (2D) roton
           is approximately by 2\,K less than the energy of a bulk (3D) roton.
           It is seen from the spectrum of 2D-  and 3D-rotons \cite{krot1,krot2,2dr} that,
            near the wall,
           3D-roton can create a 2D-roton with the emission of a 3D-phonon (a)
           or the inverse process (b) is possible.
           The creation of a 3D-roton and a 3D-phonon by a 2D-roton and the inverse process are forbidden by the energy conservation law.
           The creation of a 2D-roton must dominate over its absorption, since the former is determined only by the
           probability of the process itself, whereas the probability of the absorption is proportional else to the concentration of phonons with the required momentum.
            In other words, a 2D-roton can fuse with a 3D-phonon only if such a phonon will be near, whereas a 3D-roton can decay at once.
            At $T \gsim 1\,$K, the number of rotons is great, and
           the rotons approaching the wall must decay
           into a 3D-phonon and a 2D-roton, until
           the maximally possible concentration of 2D-rotons will be attained. In other words, $\rho_{n}$ becomes equal to
           $\rho$, and $\rho_{s}$ becomes zero on the surface, which is observed in experiments.
           At very low $T\lsim
           0.1\,$K, the number of rotons less than the number of phonons by many orders.
           In this case, if the maximally possible concentration of
           2D-rotons on the surface would be conserved as before, then process (b) would dominate, because it is proportional
            to the large concentrations of 2D-rotons and  3D-phonons, and processes (a) is proportional to a very low concentration
           of 3D-rotons.  Process (b) will continue, until, first, the concentration of 2D-rotons drops to a certain equilibrium one,
           and, second, the temperature $T$ of the wall becomes much less than $T_{\lambda}$.
           Since the spectrum of 3D-phonons at low energies coincides with the (theoretical) spectrum of 2D-phonons \cite{krot1,krot2},
             the temperatures in the bulk and on the surface must be close, i.e., $\rho_{s}\approx \rho$ at the surface.
            The critical temperature $T_{c}$, at which
            the exhaustion of $\rho_{s}$ on the wall disappears, is probably near $T_{c} \simeq 0.5 \div 1 \,$K which is
            the temperature $T$ of the transition from the dominance of rotons to the dominance of phonons.

             We note that as early as 1941 P.\,L. Kapitsa \cite{kap} observed a jump of $T$ in near-surface layers of helium near a heater.
               In this case, the heater was supplied by a heat flow. Such a jump was explained theoretically in Ref.~\onlinecite{xal-skacokT},
               but those calculations did not involved surface excitations of helium. The above-presented reasoning shows that
               a jump of $T$ must be observed in the first several atomic layers of helium near the wall due to surface excitations.
               Importantly, that in this case we have an equilibrium state without a heat flow, but with the gradient of $T$.
               The nature of this jump is different from that of the Kapitsa's jump, the latter being related to the high heat conductivity of helium
               which implies that the equilibrium is established, in the first turn, in bulk and surface helium.
                The heat exchange with the wall is much more slower.
              In this case, the heat exchange between surface excitations of helium and the wall is obviously possible.
               But it should be expected that it is insignificant. Therefore, a small jump of $T$ between the bulk wall and bulk helium
               should be present even in the absence of a heat supply to the wall or helium,
              which can be verified experimentally. In this case, it is necessary to take into account that such a jump of $T$
                 should be present between a heater and helium, as well as between a thermometer and helium.

           It is of significance that, at $T\leq T_{c}$, we must observe the ``dry'' friction,
           which can be verified in a direct experiment like \cite{gam} or by measuring the temperature
           dependence of the peak of a surface roton on the temperature
           \cite{2dr}. In experiments with the third sound
           \cite{3sound1}, it was found that the recovery length for $\rho_{s}$ increases with $T$ at $T\gsim 1\,$K
           and is constant at  $T\lsim 1\,$K. These dependencies have no explanation, and it is possible that $T \approx 1\,$K,
           at which the character of the dependence varies,
           is $T_{c}$. In Ref.~\onlinecite{gam}, the temperature
           of the experiment was not given, but such experiments are usually carried on
           at $T > 1.2\,$K. If the dry friction will not be discovered at small $T$ or
           turn out much less than that estimated in Ref.~\onlinecite{ginz1}, then
           $\rho_{s}$ is exhausted on a wall due to a decrease of the total density almost to zero, rather than due to the accumulation of 2D-rotons.

 The ideas of this section are considered in more detail in Ref.~\onlinecite{dryF}.

    \section{Electric signal for He~II with a wave of second sound}
     Let us calculate the electric signal induced by second sound in He~II at electrodes made of metals of four types
     enumerated  in Sec. III.

      \subsection{Electrodes with a strong potential} \label{4.1}

We consider electrodes made of material that solidifies the first layer of helium near
their surface.
     In the experiment \cite{rub},
      three electrodes were used: those made of gold, brass (it consists mainly of copper), and
      ruthenium oxide.  For gold and copper, the first layer of helium near the metal
      is solidified, whereas the second one is liquid with an enhanced density
      \cite{pat} corresponding to $p\simeq 10$$-$$15$~atm. As the distance to the metal
      increases by several atomic layers, the density decreases to the bulk density.

Consider the first layer of helium atoms at the electrode. This layer is polarized
perpendicularly to the electrode surface with the value of polarization given by
(\ref{3-4}) and (\ref{3-8}). To calculate $\textbf{d}_{\rm h}$, we must find $S_{7}$
(\ref{3-10}). Since the first layer is solid and the following ones are liquid but at a
varying pressure, the function $g(\textbf{r})$ in (\ref{3-10}) must be rather
complicated. However,  about $95\%$ of the contribution to $S_{7}$ comes from the nearest
layer, as is follows from our numerical calculation. Therefore, we can determine $S_{7}$ approximately, taking $g(\textbf{r})$ to
be isotropic and corresponding to the pressure of this layer. For the liquid layer, we
take $p\approx 13$~atm.  Since \cite{volpol} $S_{7}\sim \rho^{4/3}$, relation (\ref{3-8}) yields $\textbf{d}_{\rm h}\sim
\rho^{7/3}\sim \bar{R}^{-7}$. Therefore, we can determine $\textbf{d}_{\rm h}$
approximately, by multiplying (\ref{3-11}) by $\left[ 2\bar{R}_{0} / (\bar{R}_{1} +
\bar{R}_{2}) \right]^{7} \approx 1.56$, where $\bar{R}_{1}$  and $\bar{R}_{2}$ are the
mean interatomic distances in the first and second layer, respectively. Then
            \begin{equation}
           \textbf{d}_{\rm h}   \approx   -\textbf{i}_{z}\cdot 2.06 \cdot 10^{-4}|e|a_{B}.
            \label{5-1}     \end{equation}
The data on the density and thermal expansion coefficient of helium can be taken from
         Ref.~\onlinecite{es}.

We now calculate the potential induced by the first layer of helium on the electrode.
In experiments, the potential difference between the electrode on the internal surface of a resonator
 and the resonator itself (insulated from each other), whose external surface was grounded, was registered. In other words, it is
the potential difference between the electrode and a point at infinity. The dipole layer (DL) in helium near the internal
surface of the resonator induces DL of ``images'' on this surface. As a result, the potential on the internal surface of the electrode
that is induced by both DLs is equal to zero at all points. It is difficult to find the
total potential of the electrode with regard for all contributions.
But we need only the variable part of the potential difference that can be easily determined. We use the formula
\begin{equation}
           {\sss \triangle} \varphi = -\int\textbf{E}\textbf{ds}
            \label{5-2}     \end{equation}
and draw a contour from a point on the electrode surface to infinity. In calculations of the surface contribution,
we take only a part of the contour near the electrode into account. The next
part up to the resonator wall gives the bulk potential difference,
and the following part with intersections of the resonator, vessel, etc. can be omitted, since this part is invariable in
time, whereas we are interesting in the variable part of the potential created by oscillations in a wave of second sound.
From the relation $\textbf{E}+4\pi\textbf{P} \equiv \textbf{D}=0$, which holds near the resonator surface, and from (\ref{5-2}) we obtain at once the
required relation (\ref{5-7}). But it is better to obtain it also from the microscopic consideration. To this end, we note that
the potential at a point on the metal surface is zero, being the sum of potentials from DL in helium and its image in a metal.
If we move along the contour into helium, then the potential immediately after the helium DL is equal to the doubled potential of DL, because
DL and its image give now contributions of the same sign. Further along the contour, the field strength from both DLs is zero
(if we neglect the weak field from the ends of DL). Thus,
the required potential difference is equal to the potential in helium formed by two DLs. Below, we will write the potential,
keeping in mind the potential difference between the given point and the ground.

 Consider a dipole $\textbf{d}$ located in helium at the point $\textbf{r}$. At the origin of
coordinates, it creates the potential \cite{tamm}
  \begin{equation}
  \varphi = -\frac{\textbf{d}\textbf{r}}{\varepsilon r^3}.
    \label{5-6}     \end{equation}
We will determine the contribution of the first layer of helium to the potential difference between the
electrode and the ground, by summing the contributions of all microscopic
dipoles composing DL and its image:
           \begin{eqnarray}
  \varphi_{1} &=& -\sum\limits_{j}\frac{2\textbf{d}_{1}\textbf{r}_{j}}{\varepsilon r^{3}_{j}} =
    -\frac{2}{\varepsilon \bar{R}_{1||}^{2}}\int d\phi \rho d\rho \frac{d_{1,z}\times z_{0}}{(z_{0}^{2}+\rho^2)^{3/2}}=
    \nonumber \\ &=& -\frac{4\pi d_{1,z}}{\varepsilon \bar{R}_{1||}^{2}},
    \label{5-7}     \end{eqnarray}
where  $\bar{R}_{1||}$ is the mean distance (along the surface) in the first layer, and the
summation is carried out only over atoms of helium. The consideration of the mutual polarization of dipoles of first layers
varies the result only by several percents. But the formulas become awkward, and we omit these corrections.

       In the experiment \cite{rub}, a standing half-wave
       of second sound was created in the form
       \begin{equation}
       T = T_0 - 0.5 {\sss \triangle}  T(z) \cos(\omega_{2} t), \
         {\sss \triangle}  T(z) =  {\sss \triangle}  T_{0} \cos(z\pi/L_{r}).
     \label{5-8}     \end{equation}
Here,   $\omega_{2}$ is the frequency  of second sound, and $z$
   is reckoned from the resonator end in helium.
        Consider the electrode at the resonator end ($z=0$).
          Oscillations of the helium temperature lead to oscillations of
     potential (\ref{5-7}) due to the dependence of $\bar{R}_{1||}$,
     $\varepsilon,$ and $d_{1,z}$ on $T$. We now expand $\varphi_{1}$ in a power series
     with respect to $T - T_0$, leaving the first nontrivial term:
    \begin{equation}
  \varphi_{1}(T) = \varphi_{1}(T_0) + (T-T_{0})\partial\varphi_{1} /\partial T,
    \label{5-9}     \end{equation}
     \begin{eqnarray}
\frac{\partial\varphi_{1}}{\partial T} &=&
\frac{\partial\varphi_{1}}{\partial \bar{R}_{1||}}\frac{\partial\bar{R}_{1||}}{\partial T} +
\frac{\partial\varphi_{1}}{\partial \varepsilon}\frac{\partial\varepsilon}{\partial T} +\nonumber \\
&+& \frac{\partial\varphi_{1}}{\partial d_{{\rm h},z}}\frac{\partial d_{{\rm h},z}}{\partial T}
+ \frac{\partial\varphi_{1}}{\partial d_{{\rm mir},z}}\frac{\partial d_{{\rm mir},z}}{\partial T}.
\label{5-10}     \end{eqnarray}
     Then we have
        \begin{equation}
    \frac{\partial\varphi_{1}}{\partial \bar{R}_{1||}}\frac{\partial\bar{R}_{1||}}{\partial T}
    = -\frac{2}{3}\alpha^{s}_{1}\varphi_{1}.
    \label{5-11}     \end{equation}
     We denote $d_{{\rm h},z} \equiv d_{\rm h}$, $d_{{\rm mir},z} \equiv  d_{\rm mir}$,
and $d_{{\rm h},z} + d_{{\rm mir},z} = d_{1,z} \equiv d_{1}$. The quantity
      $\alpha=-n^{-1}\partial n/\partial T$ is  the thermal expansion coefficient.
        The indices $l$ and $s$ in $\alpha$ refer to
liquid and solid states, respectively, and the numerical index counts the helium layer
starting from the metal surface.

           The quantity $d_{\rm h}$ (\ref{5-1}) depends on $T$ via $\bar{R}_{1}$ and $\bar{R}_{2}$. Note that $\partial S_{7}/\partial T$
    is negligible for liquid helium \cite{volpol}. Therefore, $\frac{\partial d_{\rm h}}{\partial T} \approx
    \frac{\partial d_{\rm h}}{\partial \bar{R}_{1}}\frac{\partial \bar{R}_{1}}{\partial T}+
    \frac{\partial d_{\rm h}}{\partial \bar{R}_{2}}\frac{\partial \bar{R}_{2}}{\partial T}
     =-\frac{7d_{\rm h}}{3}\left (\frac{\alpha_{1}\bar{R}_{1}}{\bar{R}_{1}+\bar{R}_{2}}+\frac{\alpha_{2}\bar{R}_{2}}{\bar{R}_{1}+\bar{R}_{2}}\right ) $ and
     \begin{equation}
       \frac{\partial\varphi_{1}}{\partial d_{\rm h}}\frac{\partial d_{\rm h}}{\partial T}
            \approx  -\varphi_{1}\frac{7d_{\rm h}}{3d_{1}}\left (\frac{\alpha^{s}_{1}\bar{R}_{1}}{\bar{R}_{1}+\bar{R}_{2}}+
      \frac{\alpha^{l}_{2}\bar{R}_{2}}{\bar{R}_{1}+\bar{R}_{2}}\right ).
    \label{5-12}     \end{equation}

      The quantity $d_{\rm mir}$ (\ref{3-1}), (\ref{3-3}) depends on $T$ via  $z_{0}$ and $\omega_{\rm pl}$.
      We do not know exactly $\omega_{\rm pl}(T)$, but the thermal expansion coefficient
      $\alpha_{\rm m}$ for metals at helium temperatures
      is extremely small, by 5-6 orders less than that for He II \cite{nov}.
       Since the derivative
       of $\omega_{\rm pl}=\sqrt{4\pi e^{2}n_{\rm e}/m_{\rm e}}$ with respect to $T$ is proportional to
       $\alpha_{\rm m}$, it is small, and we omit it. Then $d_{\rm mir}(T)=d_{\rm mir}(z_{0}(T))$ and
            \begin{equation}
   \frac{\partial\varphi_{1}}{\partial d_{\rm mir}}\frac{\partial d_{\rm mir}}{\partial T} \approx
   - \varphi_{1} \alpha_{z} \frac{4d_{\rm mir}}{3d_{1}},
    \label{5-15b}     \end{equation}
     where $\alpha_{z}= 3\frac{\partial z_{0}}{\partial
    T}\frac{1}{z_{0}}$.
     In (\ref{5-10}), it remains to calculate $\frac{\partial\varphi}{\partial \varepsilon}\frac{\partial\varepsilon}{\partial
     T}$.   Relation (\ref{b-5}) from Appendix B yields
     \begin{equation}
  \frac{\partial\varphi_{1}}{\partial \varepsilon}\frac{\partial\varepsilon}{\partial
     T} \approx \varphi_{1}\frac{\beta_{0}\delta_{0}}{T^2} + \varphi_{1}\alpha^{l}_{1}(1- \varepsilon^{-1}).
    \label{5-20}     \end{equation}
        Near the surface metal, the DM of a helium atom consists of two parts: the stationary part
    $d_{1,z}\equiv d_{1}$ and the fluctuating one, with the mean modulus $\tilde{d}$. We note that relation
    (\ref{4-3}) includes the total DM of an atom $d_{\rm in}$, and, according to (\ref{4-3}), $\delta_{0}\sim
    d_{\rm in}^{2}$. Therefore, we need to replace $\delta_{0}$ in (\ref{5-20})
     by $\delta_{0}d^{2}_{\delta 1}/\tilde{d}^{2}$, where $d^{2}_{\delta 1}\simeq \tilde{d}^{2} +d_{1}^2$ is
     the square modulus of the total DM of a helium atom at the surface.
         As a result, we have
      \begin{equation}
  \frac{\partial\varphi_{1}}{\partial \varepsilon}\frac{\partial\varepsilon}{\partial
     T} = \varphi_{1}\frac{\beta_{0}\delta_{0}}{T^2}\frac{d_{\delta 1}^{2}(p_{1})}{\tilde{d}^{2}({\rm svp})}
     + \varphi_{1}\alpha^{l}_{1}(1- \varepsilon^{-1}).
    \label{5-23}     \end{equation}
   This formula is true, if the first layer of helium is liquid.
   If the first layer is solid, the value of
    $ \frac{\partial\varphi_{1}}{\partial
    \varepsilon}\frac{\partial\varepsilon}{\partial T}$ will be somewhat changed, but of the order of that in (\ref{5-23}).
     However, relation (\ref{5-23}) gives a small contribution to $a$ (\ref{new5}),
    and we omit this correction in what follows.
     Finally, potential (\ref{5-9}), (\ref{5-10}) takes the form
          \begin{eqnarray}
          \frac{\partial\varphi_{1}}{\partial T} &\approx &
  - \frac{7d_{\rm h}}{3d_{1}}\left (\frac{\alpha^{s}_{1}\bar{R}_{1}}{\bar{R}_{1}+\bar{R}_{2}}+
      \frac{\alpha^{l}_{2}\bar{R}_{2}}{\bar{R}_{1}+\bar{R}_{2}}\right )\varphi_{1} - \nonumber \\
   &-& \frac{2}{3}\alpha^{s}_{1}\varphi_{1} - \frac{4d_{\rm mir}}{3d_{1}}\alpha_{z}\varphi_{1}.
        \label{5-26a}     \end{eqnarray}
           The quantity
    $\alpha_{z}= 3z_{0}^{-1}\partial z_{0}/\partial
    T$ is determined by the
    distance $z_{0}$ from the  helium atoms to the metal.
    The value of $\alpha_{z}$ can be estimated with the help of the relation $\alpha_{z}\sim
    \alpha D_{\rm h}/D_{\rm m}$, where  $D_{\rm h}\approx 11\,$K is the depth
    of the interaction potential between two $^4$He atoms,
    and $D_{\rm m}$ is the depth of the $^4$He atom--metal potential.
    The estimate follows from the relation $ \delta \bar{R}\sim \delta V\sim
    D$, where $V$ is the corresponding potential. In the case where the first layer of helium is solid, we have $\alpha_{z}\sim
    \alpha^{s}_{1}D_{\rm h}/D_{\rm m}$, and
           \begin{eqnarray}
            \frac{\partial\varphi_{1}}{\partial T} &\simeq &
         -\alpha^{l}_{2}\varphi_{1}\frac{7d_{\rm h}}{3d_{1}}\frac{\bar{R}_{2}}{\bar{R}_{1}+\bar{R}_{2}} - \label{5-26}  \\
           & -& \alpha^{s}_{1}\varphi_{1}\left ( \frac{2}{3} +
   \frac{4d_{\rm mir}}{3d_{1}}\frac{D_{\rm h}}{D_{\rm m}}+
   \frac{7d_{\rm h}}{3d_{1}}\frac{\bar{R}_{1}}{\bar{R}_{1}+\bar{R}_{2}} \right ).
      \nonumber     \end{eqnarray}
     In the presence of the wave of second sound (\ref{5-8}), the potential $\varphi_{1}(T)$ oscillates with the frequency of second sound
     and with the amplitude
          \begin{equation}
   {\sss \triangle} \varphi_{1}(z)  = \frac{\partial\varphi_{1}}{\partial T}(T_0) {\sss \triangle} T(z).
    \label{5-25}     \end{equation}
   It is convenient to represent $  {\sss \triangle} \varphi_{1}$ in the form
    \begin{equation}
    {\sss \triangle} \varphi_{1}(z) = \frac{k_{B}
    {\sss \triangle} T(z)}{2|e|} a_{1},
    \label{5-27}     \end{equation}
 \begin{eqnarray}
  a_{1}  \approx \frac{5.78K n^{2/3}_{1}}{\varepsilon 10^{-4}|e|a_{B}n^{2/3}_{0}}
    \left \{\frac{7\alpha^{l}_{2}d_{\rm h}}{1+\bar{R}_{1}/\bar{R}_{2} } \right.+ \nonumber \\
   + \left. \alpha^{s}_{1}\left (2d_{1} + d_{\rm  mir}\frac{4D_{\rm h}}{D_{\rm m}} +  \frac{7d_{\rm h}}{1+\bar{R}_{2}/\bar{R}_{1} }\right )
      \right \}.
         \label{5-28}     \end{eqnarray}

         Now let us consider the second and subsequent layers of helium.  In helium near the wall,
the pressure $p$ at the very wall is maximum (25~atm), but it drops to the bulk pressure
as the distance from the wall increases by  2--3 atomic layers \cite{pat}. Due to the
difference in pressures, the concentration of helium atoms in the first and third layers
are different. Therefore, the tidal actions of these layers on the second layer are also
different. As a result, the second layer is polarized. The third layer becomes polarized
as well.  But the polarization of the fourth layer is negligible, since the pressure in the third layer
coincides almost with the bulk one \cite{pat}.  According to formula (20) in
Ref.~\onlinecite{volpol}, the polarization of the
           $j^{\rm th}$ layer is given by
            \begin{eqnarray}
             \textbf{d}_{j\geq 2} &\approx &
   d_{0}\textbf{i}_{z}\frac{n_{j-1}S_{7}(n_{j-1})-n_{j+1}S_{7}(n_{j+1})}{4n_{0}} =\label{new1}  \\
   &=& d_{0}\textbf{i}_{z}\frac{S_{7}(n_{0})}{4}\left \{\left (\frac{n_{j-1}}{n_{0}}\right )^{7/3}
    -\left (\frac{n_{j+1}}{n_{0}}\right )^{7/3}\right \}.
                    \nonumber  \end{eqnarray}
(we note a small mistake in Ref.~\onlinecite{volpol}: the sign of $d_{0}$ in formula (11)
is wrong; after its correction, the signs in the subsequent formulas become correct,  and
$d_{0}$ will have the sign in accordance with formula (\ref{3-9}) of the present work).
The polarization (\ref{3-1}) induced by the metal is negligible for these layers.
Similarly to the calculation in the case of the first layer, we obtain
     \begin{equation}
  \varphi_{j\geq 2}  = -\frac{4\pi d_{j}}{\varepsilon \bar{R}_{j||}^{2}},
    \label{new2}     \end{equation}
        \begin{eqnarray}
    a_{j\geq 2}  &\approx & \frac{53.63K\cdot n^{2/3}_{j}}{\varepsilon n_{0}^{3}}\left
    \{ n^{7/3}_{j-1}\left (\frac{2}{7}\alpha_{j}+ \alpha_{j-1}\right ) \right. - \nonumber \\
   &-& \left. n^{7/3}_{j+1}\left (\frac{2}{7}\alpha_{j}+ \alpha_{j+1}\right)  \right \},
    \label{new3}     \end{eqnarray}
    where  $\alpha_{j}\equiv \alpha_{j}(T_{j},p_{j})$.
         By considering all layers, we get
          \begin{equation}
  \varphi = \sum\limits_{j=1}^{\infty}\varphi_{j},
    \label{new4}     \end{equation}
       \begin{equation}
        {\sss \triangle} \varphi  = \frac{k_{B}  {\sss \triangle} T}{2|e|}
       a, \quad a = \sum\limits_{j=1}^{\infty}a_{j}.
                \label{new5}     \end{equation}
In our case, $a_{j\geq 4}$ are small and give a contribution to the bulk $ {\sss \triangle}
\varphi$ calculated in Ref.~\onlinecite{volpol}.
        Therefore,
         \begin{equation}
       a = a_{1}+a_{2}+a_{3}.
                \label{new6}     \end{equation}

\begin{figure}[ht]
\centerline{\includegraphics[width=85mm]{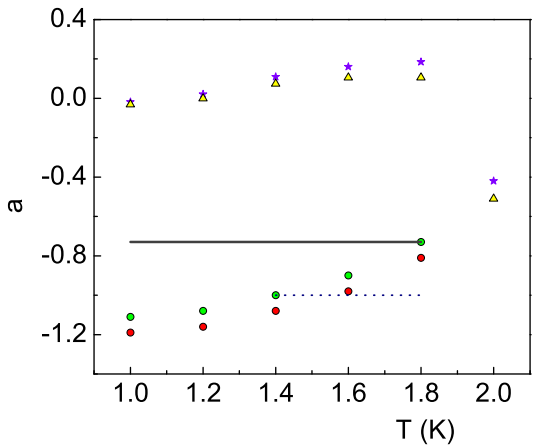}
}
\caption{Theoretical values of
$a(T)$ determining the potential difference $ {\sss \triangle} U
= a k_{B}  {\sss \triangle} T/2|e| $  between an electrode at the
resonator end and the ground: (a)~without taking into account the influence of the
metal on the temperature of the near-surface layers of helium: $\star$ denotes the quantity $a(T)$ for Au,
and ${\sss\triangle} $  for Cu; (b)~taking into account the influence of the metal on the temperature of the first three
near-surface layers of helium: $\circ$ corresponds to Au, and $\bullet$ to Cu;
   (c)~taking into account the influence of the metal on the temperature of the first \textit{four}
near-surface layers of helium: the solid line shows $a(T)$ for Au. The dashed line is
the experimental dependence $a(T)$ for Au and brass. \label{fig1}}
\end{figure}

 \begin{table}
     \begin{center}
\caption{The parameters of metals and the dipole moments ($d_{j}$) of helium atoms in
three first layers at the metal surface; the notation is given in the text. The values of
$d_{\rm h}$ and $d_{j}$ for Au and Cu are given for $p_{2}= 13$~atm. \label{table1}}
\vskip 10mm
     \begin{tabular}{|c|c|c|c|c|c|}  \hline
                        & Au   & Cu      & Cs & Na   \\ \hline
   $z_{0},\mbox{\AA}$  & 3.17 & 3.59  &  5.7 &  5.31 \\ \hline
   $D_{\rm m}$, K       & 92.8 & 59.0  & 7.0 & 12.5   \\ \hline
  $\hbar\omega_{\rm pl}$, \mbox{eV}& 25.8 & 20  & 3.3 & 5.78  \\ \hline
  $\frac{d_{\rm  mir}}{10^{-4}|e|a_{B}}$ & 2.92& 1.52  & 0.056 & 0.12  \\ \hline
  $\frac{d_{\rm h}}{10^{-4}|e|a_{B}} $  & -2.06& -2.06  & -1.32 & -1.32  \\ \hline
  $\frac{d_{1}}{10^{-4}|e|a_{B}} $  & 0.85& -0.54  & -1.26 & -1.2  \\ \hline
  $\frac{d_{2}}{10^{-4}|e|a_{B}} $  & 1.16 & 1.16  & $\approx 0$ & $\approx 0$  \\ \hline
  $\frac{d_{3}}{10^{-4}|e|a_{B}} $  & 0.41 & 0.41  & 0 & 0  \\ \hline
               \end{tabular} \end{center} \end{table}
          We now determine the quantity $a$ for electrodes made of Au and  Cu,
          located at the resonator end ($z=0$).
          In Table~\ref{table1}, we present the known parameters of
     metals $\hbar\omega_{\rm pl}$  \cite{m2,dav},  $D_{\rm m},$ and $z_{0}$ \cite{zar1,zar2},
       and those determined in the present work:
     $d_{\rm  mir}$,  $d_{1}$,  $d_{2},$  and $d_{3}$.
     If the temperature $T$ in all layers coincides with the bulk temperature of He~II,
        the value of $a$ is an order of magnitude smaller than the experimental one, has opposite sign (the sign of $a$
      determines the signal polarity), and increases with $T$ (see Fig.~\ref{fig1}).

         The experiment shows that $\rho_{s}=0$ at the wall. Since
    helium wets the wall, but $\rho_{s}$
     increases to the bulk value only at a distance of 2--3 atomic
       layers \cite{pat} from the wall, the exhaustion of $\rho_{s}$ must be related not to the smallness of the
       total density, but to the behavior of quasiparticles at the wall
        (see Sec. III and, in more detail, Ref.~\onlinecite{dryF}).
               The equality of $\rho_{s}$ to zero is the condition for the $\lambda$-transition.
          Therefore, $T=T_{\lambda}$ at the wall, and $T$ decreases smoothly to the bulk value as the distance to the wall increases.
      Note that the distance between rotons at $T \simeq T_{\lambda}$ is about two interatomic distances,
      i.e., it is approximately equal to the roton size ($\sim 3$ atomic layers, according to the experimental
          scattering cross-section of rotons by vortex lines \cite{rr}).
          Hence, the free path of a roton does not exceed the interatomic distance. Therefore, we may consider
        the temperatures of separate atomic layers.
            If the temperature of helium is considerably smaller than $T_{\lambda}$,
            then the temperature of the first layers is determined
       by surface quasiparticles.
       As the bulk temperature is changed by $\delta T$, the temperature of the $j^{\rm th}$ layer is changed by
       $\delta T_{j} \approx \delta T$. In the estimates, we use a linear law
       $T_{j}=T_{\lambda}(p_{j})-j\cdot 0.1\,$K\@.
       If the pressure and the temperature are equal to the bulk values, by starting from the 3$^{\rm rd}$ and
          4$^{\rm th}$  (the 3$^{\rm rd}$ liquid one) layers, respectively, we obtain
           $p_{1} \gtrsim 30$~atm, $p_{2}\approx 13$~atm, $p_{j\geq 3} =
       {\rm svp}$, $n_{1}/n_{0}\approx 1.31$, $n_{2}/n_{0}\approx 1.123$, $n_{j>2}/n_{0}=1$,
       $T_{1}\approx 1.8\,{\rm K}$, $T_{2}\approx T_{3} \approx
       2\,{\rm K}$, $T_{j\geq 4}= T$. We note that $T_{\lambda}$ depends on $p$.
       For $\alpha_{1}\equiv \alpha^{s},$ we used the data for the hcp phase \cite{es} (solid helium can be present also
       in the bcc phase, but the interval of relevant pressures is very narrow).
           The resulting $a$ (see Fig.~\ref{fig1}) for Au and Cu are close to the experimental values $a_{\rm exp} \approx -1$ and increase with the temperature.
            The experiment indicates that $a$ does not depend on $T$. Theoretically, this is obtained in the case where the temperature of
            the 4$^{\rm th}$ layer is also determined by the wall ($\approx 1.8\,{\rm K}$).  Then $a_{\rm Au} \approx -0.73$
            (the solid line in Fig. 1) and $a_{\rm Cu} \approx -0.81$.

                  Thus, with regard for the exhaustion of $\rho_{s},$ the theoretical value of $a$ corresponds approximately
         to the experimental one  and does not depend on the temperature.

         We note that, in the mode where the temperature of the first four layers is determined by the wall, we have
         $a_{1} \approx 3.05$, $a_{2} \approx 0.25$, $a_{3} \approx -4.03,$ and $a=a_{1}+a_{2}+a_{3} \approx -0.73$ for Au.
         As we can see, the first and third layers give the main contribution, but with different signs.
         This determines the sensitivity of $a$ to the value of $p_{2}$. The analysis indicates that,
           at $p_{2} = 10$~atm, we obtain $a\approx 0.05$ for Au. For smaller values of $p_{2}$, the quantity
         $a$ increases and reaches $\sim 1,$ which corresponds to the experimental value in magnitude
         but has opposite polarity.
          At $p_{2} \gtrsim 12$~atm, we have $a \sim  -1$, i.e., the signal agrees with the
          experimental one in magnitude and polarity.

      We do not consider the case where two or more solid atomic layers of helium are present near the metal,
        since no such metals are reliably known.

            \subsection{Electrodes with a low potential} \label{4.2}

\begin{figure}[ht]
\centerline{\includegraphics[width=85mm]{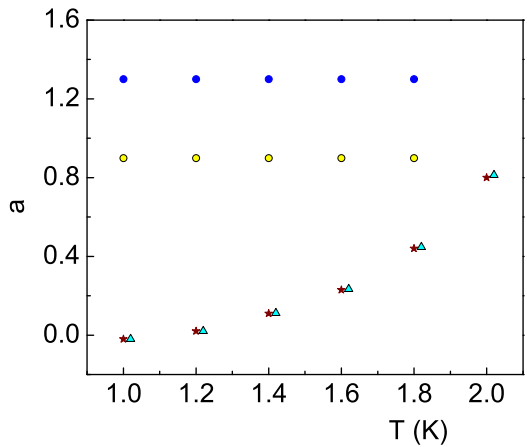}
}
\caption{The same as in Fig.~\ref{fig1}. (a)~Without taking into account
the influence of a metal on the temperature
   of the near-surface layers of helium: the quantity $a(T)$ is given for Na ($\star$) and
    Cs ($ {\sss \triangle}  $); (b) taking into account this influence: $\circ$  corresponds to Na, and $\bullet$ to
    Cs. \label{fig2}}
\end{figure}

         (\textsl{i}) Let helium wet a metal without solidifying at its surface.
                This is true for
       alkaline metals Rb, Na, K, and Li. The analysis is similar to the group of Sec.~\ref{4.1}. The difference consists in that
       the first layer of helium is liquid, rather than solid. For these metals, the helium atom--metal potential is close to the
       $^4$He--$^4$He potential. Hence, the metal does not attract strongly helium atoms, and the pressure in the first layers
       of helium near the metal is close to the bulk one (svp).
       Therefore,  only the first layer of helium is polarized. For the rest of layers,
       we have $\nabla n =0,$ and polarization is absent, according to (\ref{new1}).
       Relation (\ref{5-28}) with $\bar{R}_{2}=\bar{R}_{1}=\bar{R}_{0}$ and $\alpha_{1}^{s} \rightarrow \alpha_{1}^{l}$, yields
           \begin{eqnarray}
           a_{1}  &\approx & \frac{5.78K}{\varepsilon 10^{-4}|e|a_{B}}
    \left \{\frac{7}{2}\alpha^{l}_{2}d_{\rm h} + \right. \nonumber \\
    &+& \left. \alpha^{l}_{1}\left (2d_{1} + d_{\rm  mir}\frac{4D_{\rm h}}{D_{\rm m}} +  \frac{7}{2}d_{\rm h}\right )
      \right \}
            \label{5-33}     \end{eqnarray}
      with $\textbf{d}_{\rm h}$  given by (\ref{3-11}).  The resulting $a$ for Na are presented in Fig.~\ref{fig2}.
         We have considered two  cases: (1)~$T_{1}$ and $T_{2}$ are equal to
       the bulk temperature, (2)~$T_{1}$ and $T_{2}$ are determined by the wall ($T_{1}\approx T_{\lambda}({\rm svp}) - 0.1\,{\rm K} \approx
         2.07\,$K,  $T_{2}\approx T_{1}- 0.1\,$K). In the first case, $a < 1$ and rapidly increases with the temperature;
         in the second case, we have $a\approx 0.9$.
       The value of $a$ for K, Li, and Rb is almost the same as that for Na since the
       difference between these values is connected with $d_{\rm  mir}$, which is very small for these metals.

      (\textsl{ii}) Consider a metal that is not wetted by helium.
         Such a metal \cite{smach} is Cs.
        Due to the nonwettability, the He~II surface is separated from the metal surface by the distance $z_{0}$.
         To our knowledge, the question about the vanishing of $\rho_{s}$ at the free surface of He~II remains to be open.
         The surface rotons are not ``glued'' to a metal, but they
       propagate along the free surface
       of He~II\@. However, the dispersion curve of surface rotons should be close to that for the case where
       He~II wets the metal. Therefore, the reasoning of Sec. III and
       Ref.~\onlinecite{dryF} implies that $\rho_{s}$ must be exhausted at
       the free surface: $\rho_{s}=0$ and $T=T_{\lambda}$.

     The quantities $\textbf{d}_{\rm h}$ and $a$ are determined by formulas (\ref{3-11}) and (\ref{5-33}).
      We calculated $a$ for the same two cases as in item (i). In the case where
      $T_{1}$ and $T_{2}$ are determined by the wall, we need to take into account that $T_{\lambda}$ is attained at the surface
      of helium. Due to zero oscillations, the atoms are located not on the surface, but at a distance
      $\sim 1\,\mbox{\AA}$, on the average, below it. Respectively,
      we have $T_{1}\approx T_{\lambda}({\rm svp}) - 0.1\,{\rm K}\mbox{\AA}/\bar{R}_{0} \approx
         2.14\,$K,  $T_{2}\approx T_{1}- 0.1\,{\rm K} \approx 2.04\,$K.
       The resulting $a$
               is presented in Fig.~\ref{fig2}. As one can see, the points coincide practically with those for Na in case (1).
       In case (2), we have $a\approx 1.3$, which is by a factor of 1.4 greater
       than the same quantity for Na and other alkaline metals.
        The possibility for Cs to possess a special value of $a$ has been already discussed
       in Ref.~\onlinecite{gut}.

          The polarity $S$
           of a signal is determined by the signs of $\varphi$  and $\partial\varphi/\partial T$.
             Let us denote the experimental $S$ by $S=(+)$. Then, according to the model,
        all metals of group~A, i.e., Au and Cu, have $S=(+)$, whereas the alkaline metals
        (Na,  Cs, and others) have $S=(-)$.

\section{The bulk polarization of helium induced by the surface dipole layer.}
 In work \cite{volpol}, formula (37) presents the bulk polarization of helium arisen in a wave of second sound due to the gradients of the density
(main contribution) and the temperature:
         \begin{equation}
  \textbf{P}(Z) = n\textbf{d}(Z)/\varepsilon \approx 3.5S_{7}d_{0}n\bar{R}\nabla T(Z) \alpha/3\varepsilon.
        \label{vp-1}     \end{equation}
In Ref.~\onlinecite{volpol}, the polarization was determined without $\varepsilon$ in the denominator, and $\varepsilon$ was considered in the potential.
But more accurate to consider $\varepsilon$ namely in $\textbf{P}$, because $\textbf{d}$ is the proper DM of an atom and is screened by the medium.
Such a polarization induces the electric signal on the electrode that is weaker by one-two orders of magnitude than
the surface signal determined in the previous section.
However, there exists one more source of the bulk polarization that was not considered in Ref.~\onlinecite{volpol}. It is the above-considered surface DL (dipole layer).

\begin{figure}[ht]
\centerline{\includegraphics[width=85mm]{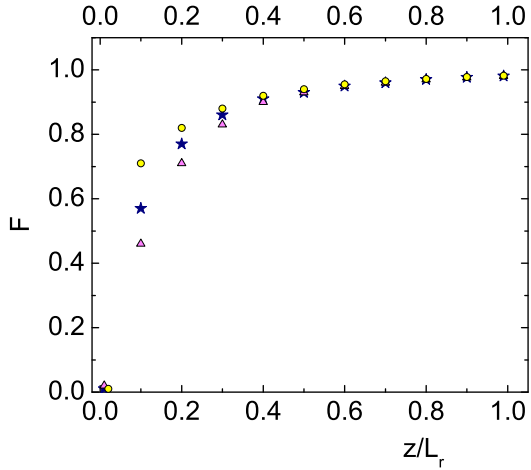}
}
\caption{Dependence of the function
$F=zf(z)(z^{2}+R^{2}_{r})^{-1/2}$ on $z$ at $\sigma = 0.2$ and
several values of $R$: $R\leq 0.2 R_{r}$
($\triangle$), $R=0.7 R_{r}$
($\star$), and $R=0.99 R_{r}$ ($\bullet$). For points
on the resonator axis ($R=0$), we have $f=1$; the corresponding
$F(z)$ coincides with the curve $\triangle\triangle\triangle$.
 At $z\gsim 0.5L_{r}$, three curves merge to the one curve. \label{fig0}}
\end{figure}

According to the analysis in the previous section, DL consisting of
several strongly polarized layers of helium atoms is formed on all internal surfaces of the resonator.
Values of DM of an atom in three first layers are given in Table 1. Let us consider two inner end surfaces of a cylindrical
resonator. As known, the infinite DL
creates no field $\textbf{E}$ outside of itself. But the real layer is a thin disk with finite radius equal to the
resonator radius $R_{r}$. Such a layer creates the field outside of itself as well. The first layer of DL creates the potential
\begin{eqnarray}
  \varphi_{1}(\textbf{R}) &=& \sum\limits_{j}\frac{2\textbf{d}_{1}(\textbf{R}-\textbf{r}_{j})}{\varepsilon |\textbf{r}_{j}-\textbf{R}|^{3}} =
    \frac{2n_{2,1}d_{1}Z}{\varepsilon }\int\limits_{0}^{2\pi} d\phi \times \nonumber \\
     &\times& \int\limits_{0}^{R_{r}} \frac{\rho d\rho}{(Z^{2}+\rho^2 +R^{2}-2R\rho\cos{\phi})^{3/2}}=
    \nonumber \\ &=& \frac{4\pi n_{2,1}d_{1}}{\varepsilon}\left (1-\frac{Z f(Z,R)}{\sqrt{Z^{2}+R^{2}_{r}}} \right ),
    \label{vp-2}     \end{eqnarray}
in helium at a point $\textbf{R}=(R, 0, Z).$ Here, $\textbf{R}_{j}=(\rho, \phi, 0)$ are coordinates of
the $j^{\textrm{th}}$ atom of helium in the layer, $n_{2,l}$ is the surface concentration of helium atoms
in the $l^{\textrm{th}}$ layer, and
\begin{eqnarray}
  f(Z,R) &=&\frac{\sqrt{\tilde{Z}^{2}+1}}{\pi} \int\limits_{0}^{\pi} d\phi\frac{ (\tilde{Z}^{2}+\tilde{\rho}^{2} +\tilde{\rho}\cos{\phi})}{(\tilde{Z}^{2}+\tilde{\rho}^{2}\sin^{2}{\phi})}\times \nonumber \\
  &\times&(\tilde{Z}^{2}+\tilde{\rho}^{2} +1+2\tilde{\rho}\cos{\phi})^{-1/2},
        \label{vp-3}     \end{eqnarray}
where $\tilde{Z}=Z/R_{r}$, $\tilde{\rho}=R/R_{r}$. For the points on the resonator axis, we have $f=1$. In Fig.~\ref{fig0}, we show the numerically determined
dependence of the function $Zf(Z,R)(Z^{2}+R^{2}_{r})^{-1/2}$ on $Z$ at several $R,$ by comparing it with the same function at $f=1$.
The exact factor $f$ flattens the dependence on $Z$ and weakens the strength $E_{z}$. We note that formulas
(\ref{vp-2}) and (\ref{vp-3}) do not involve the contribution of images of the dipoles arising on the electrode and walls of the resonator.  Below, we will
use formula (\ref{vp-3}) with $f=1$ for estimates. Then the integrals can be calculated analytically.

Considering that DL is present on both ends of the resonator and consists of three atomic layers and taking $f=1,$ we obtain the total potential:
 \begin{equation}
\varphi(\textbf{R})\approx  \frac{4\pi n_{2,ef}d_{ef}}{\varepsilon}f_{2}(Z)=
 -\varphi_{s}f_{2}(Z),
        \label{vp-4}     \end{equation}
\begin{equation}
f_{2}(Z)=\left (2-\frac{Z}{\sqrt{Z^{2}+R^{2}_{r}}} -
\frac{L_{r}-Z}{\sqrt{(L_{r}-Z)^{2}+R^{2}_{r}}} \right ).
        \label{vp-4b}     \end{equation}
Here, $\varphi_{s}$ is the surface potential on the electrode from three layers given by (\ref{new5}), and
$n_{2,ef}d_{ef}=n_{2,1}d_{1}+n_{2,2}d_{2}+n_{2,3}d_{3}$. Potential (\ref{vp-4}) creates the field
$\textbf{E}=-\textbf{i}_{z}\partial\varphi/\partial Z-\textbf{i}_{\rho}\partial\varphi/\partial R$.
If we consider the exact value of $f,$ the analysis indicates that the $E_{\rho}$ component is of the order of magnitude of the $z$-component but is somewhat less.
In the approximation where $f=1,$ we have $E_{\rho}=0$, and the $z$-component, $E_{z}$, induces the bulk polarization
\begin{eqnarray}
 &&\frac{nd_{z}}{\varepsilon}= P_{z} = \frac{(\varepsilon - 1)E_{z}}{4\pi} =   \frac{(\varepsilon - 1)\varphi_{s}}{4\pi }
 \times \nonumber \\
 &\times & \left (\frac{R^{2}_{r}}{\sqrt{Z^{2}+R^{2}_{r}}} -
\frac{R^{2}_{r}}{\sqrt{(L_{r}-Z)^{2}+R^{2}_{r}}} \right )
        \label{vp-5}     \end{eqnarray}
in helium. As was noted above, taking into account the exact function $f$ causes a decrease in both $E_{z}$ and the polarization. On the end electrode ($Z=0$), polarization (\ref{vp-5}) induces
the potential
 \begin{equation}
  \varphi_{bs} \approx -\sum\limits_{j}\frac{\textbf{d}(\textbf{r}_{j})\textbf{r}_{j}}{\varepsilon r_{j}^{3}} =  \frac{(\varepsilon - 1)\varphi_{s}I_{bs}}{2},
        \label{vp-6}     \end{equation}
  \begin{eqnarray}
  I_{bs} &=& \sigma^{2}\int\limits_{0}^{1} dz \left (1-\frac{z}{\sqrt{z^{2}+\sigma^{2}}} \right )\times\nonumber \\
  &\times &\left (\frac{1}{(z^{2}+\sigma^{2})^{3/2}}- \frac{1}{((1-z)^{2}+\sigma^{2})^{3/2}} \right ),
        \label{vp-7a}     \end{eqnarray}
where $\sigma = R_{r}/L_{r}\lsim 1$. However, one need to take into account that we are interested in the variable part of the potential.
Therefore, one should change the sign of the contribution of the second resonator end (in Eqs. (\ref{vp-5}) and (\ref{vp-7a})), because the value of $\delta T$ has different sign at different resonator ends, in the half-wave of second sound.
So, we obtain for $I_{bs}$ in Eq. (\ref{vp-6}):
 \begin{eqnarray}
  I_{bs} &=& \sigma^{2}\int\limits_{0}^{1} dz \left (1-\frac{z}{\sqrt{z^{2}+\sigma^{2}}} \right )\times\nonumber \\
  &\times &\left (\frac{1}{(z^{2}+\sigma^{2})^{3/2}}+ \frac{1}{((1-z)^{2}+\sigma^{2})^{3/2}} \right ).
        \label{vp-7}     \end{eqnarray}
 The numerical calculation for (\ref{vp-7}) gives $I_{bs} \approx 0.38; 0.7$ at $\sigma = 1/24; 1/2$ ($\sigma$ in first experiment),
$I_{bs} \approx 0.51$ at $\sigma = 0.2$ (new experiment). As a result,
the bulk potential $\varphi_{bs}$ is equal to the above-calculated surface potential $\varphi_{s}$ multiplied by a small factor
($\lsim 0.02$) that depends on the ratio of resonator sizes.
Such a potential leads to a small correction to signal (\ref{new5}), in the limits $\sim 2\%$. This correction is larger for short resonators,
as for polarization (\ref{vp-1}).

Polarization (\ref{vp-5}) is much greater than polarization
(\ref{vp-1}). But the main part of (\ref{vp-5}) is invariable in
time and unobservable. The variable part is, on the average, by one
order of magnitude less than polarization (\ref{vp-1}),
their ratio $\sim -0.1a \alpha(1.8K)/\alpha(T)$. In this case,
the strength of the field induced by DL turns out, on the average,
approximately by one order of magnitude more than the strength induced by a
wave of the spontaneous polarization (\ref{vp-1}).

Besides ends, DL is present also on the lateral surfaces of a resonator. We omit the calculation of the potential of this DL, since a lateral
surface can be divided into segments and represent in the form of a fan of pairs of almost plane surfaces positioned oppositely to each other.
By this, the problem is reduced to the previous one, and, hence, the signal should be of the same order, i.e., it should be low.
The exact calculation without regard for the above approximations is complicated and can be performed only numerically. However, the above-found estimate
must give the true order of magnitude.

Thus, DL on the internal surface of a resonator induces the polarization in the bulk of helium that is variable in time and
approximately by one order of magnitude less than spontaneous polarization (\ref{vp-1}) related to the density gradient. The distribution of the induced polarization   is such that the electric
signal from it turns out low, $\sim 1\%$ of the surface signal. Though, for a long resonator from the first experiment \cite{rub},
this signal is of order of magnitude of the signal from polarization (\ref{vp-1}) equal to \cite{volpol} $\sim 0.1-1\%$ of the surface signal.
Thus, we will neglect the bulk signal (\ref{vp-6}), since it gives only a small correction to the surface signal.

Note that the potential (\ref{vp-4}) with opposite sign represents the more exact formula for the surface signal, as compared with (\ref{new5}), because it considers the
contribution of the remote resonator end. Consideration of the DL on a lateral walls of the resonator will lead to an additional correction. All these corrections
represent,  in essence, a bulk ones. Due to them, the surface signal acquires a weak dependence on the resonator sizes, which is apparent from Eqs. (\ref{vp-4}) and (\ref{vp-4b}). Here, we do not consider a possible difference between materials of the resonator ends. It is of interest that a weak dependence of the signal on the resonator sizes should be even at purely surface nature of the signal.

 \section{Influence of admixtures on the signal induced by second sound.}
 We now estimate the influence of admixtures on the polarization of helium and the amplitude of the electric signal. This question was not studied earlier.

 First, we consider an admixture of nonpolar molecules or atoms.
 As an example, we consider the admixture of $^3$He atoms, since their properties are studied quite well.
    The equations for the sounds for $^3$He--$^4$He mixtures are modified \cite{xal}.
  A change of the temperature of the mixture by
$ {\sss \triangle}  T$ leads to a change in the concentration of $^3$He atoms, according to
equations (54.6) and  (57.3) from Ref.~\onlinecite{xal}:
     \begin{equation}
         {\sss \triangle}  f  = \frac{f {\sss \triangle}  T \times\partial S_{\rm mix}/\partial T + f {\sss \triangle}  p
        \times\partial S_{\rm mix}/\partial p}{S_{\rm mix}-f\times\partial S_{\rm mix}/\partial f},
         \label{4-1}     \end{equation}
     \begin{equation}
        S_{\rm mix}  = S_{0}+k_{B}n_{4}f\ln{\left [\frac{2}{n_{4}f}\left (\frac{m^{*}k_{B}T}{2\pi\hbar^{2}}  \right )^{3/2}+\frac{5}{2}\right ]},
    \label{4-2}     \end{equation}
     \begin{equation}
       f= \frac{n_{3}m_{3}}{n_{3}m_{3}+n_{4}m_{4}},
    \label{4-2b}     \end{equation}
          where $S_{\rm mix}$ is the entropy of the mixture per unit volume, $S_{0}$ is the entropy of pure $^4$He,
           $n_{3}$ and  $n_{4}$ are the concentrations of $^3$He and $^4$He atoms,
           respectively, and
          $m^{*}\approx m_{3}+m_{4}$ is the effective mass of the He atom in the mixture.
      For the second sound, we have
       \begin{equation}
         {\sss \triangle}  f  \approx f {\sss \triangle}  T \frac{1}{S_{0}}\frac{\partial S_{0}}{\partial T},
            \label{4-3}     \end{equation}
       hence, the full concentration, equal to $n_{\rm mix}=n_{3}+n_{4}\approx \left[ 1+4f(3-3f)^{-1} \right]n_{4}$,
        acquires an additional contribution from $ {\sss \triangle}  f$:
        \begin{equation}
         {\sss \triangle}  n_{\rm mix}  \approx \frac{4f}{3(1-f)^2}n_{4} {\sss \triangle}  T \frac{1}{S_{0}}\frac{\partial S_{0}}{\partial T}.
            \label{4-4}     \end{equation}
           At $T=1.4$$-$$2$~K  and $f=0.1$, we obtain $ {\sss \triangle}  n_{\rm mix}\simeq 0.4n_{\rm mix} {\sss \triangle}  T/{\rm K}$, which is two orders of magnitude larger than
         the quantity
          $ {\sss \triangle}  n_{4}\approx -n_{4}\alpha  {\sss \triangle}  T$  for the second sound in pure $^4$He. If $^3$He atoms do not form pairs
          with $^4$He (the idea of such pairing is proposed in Ref.~\onlinecite{fvp}), then the polarizational properties of $^3$He and $^4$He atoms should
          be similar. In this case, an admixture of $^3$He atoms will lead to an additional signal $U$,
          proportional to $f$.  This signal consists of the (main) surface part and the volume part, as for the pure $^4$He.
          We can estimate the surface part by multiplying the signal (\ref{new5}) for the pure $^4$He  by the ratio
           $ {\sss \triangle}  n_{\rm mix}$  to  $ {\sss \triangle}  n_{4}\approx -n_{4}\alpha  {\sss \triangle}  T$,
           all taken at the temperature of the first layer of the helium  near the electrode ($T\simeq 2\,$K).
      We obtain that, at high concentration of $^3$He ($f=0.1$), the additional signal in the wave of the second sound, for the golden electrode,
      must be \textit{one or two orders of magnitude larger} than the already observed signal $ {\sss \triangle}  U  \approx -k_B  {\sss \triangle}  T/2|e|$.
      If $^3$He atoms pair with $^4$He, then the effect should probably be stronger,
   since such molecules are  electrically more active then the spherically symmetric $^3$He atoms. Thus, the random
admixtures of nonpolar molecules do not affect the properties of a signal at low concentrations, but they can significantly enhance both
surface and bulk signals at high concentrations.

\begin{figure}[ht]
\centerline{\includegraphics[width=85mm]{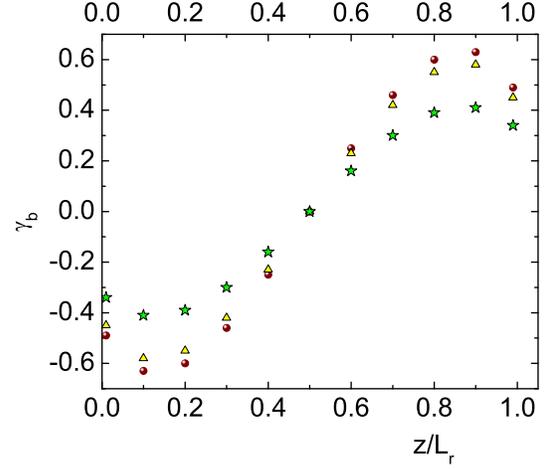}
}
\caption{Dependence of the quantity $\gamma_{b}$ (\ref{4-9}) on $z$ at $\sigma = 0.2$ and $R=0$
($\bullet$), $R=R_{r}/2$ ($\triangle$), $R=R_{r}$ ($\star$).  \label{fig3}}
\end{figure}

We now carry on several estimates for admixtures of polar molecules. Let the molecules be located in the bulk of He II. The electric field $\textbf{E}$
induced by atoms of helium will put the proper DMs of molecules in order. So, the DMs will be so oriented to decrease the
external field. Let us estimate this effect.
The mean projection of the proper DM $\textbf{d}_{p}$ of a molecule on the external field $\textbf{E}$ is determined by the well-known formula
 \begin{equation}
  d_{E}=\frac{d_{p}^{2}E}{3k_{B}T}.
        \label{4-0}     \end{equation}
In the bulk, the field $\textbf{E}$ is different at different points and is directed mainly along the $Z$ axis of a resonator. A method of calculation
of the electric field in a spontaneously polarized dielectric is discussed below in Appendix A.
According to this method, the potential at a point with coordinates $\textbf{R}=(R, 0, Z)$ can be determined as in Refs.~\onlinecite{lt1,volpol}, by summing
the potentials from all local dipoles in the resonator:
         \begin{eqnarray}
   & \varphi_{b}(\textbf{R}) &= n\int d\textbf{r}\frac{\textbf{d}(z)(\textbf{R}-\textbf{r})}{\varepsilon
   |\textbf{r}-\textbf{R}|^{3}}= \label{4-5}  \\   &=&
     \int dV\frac{nd_{z}(z)(Z-z)}{\varepsilon
   [(z-Z)^{2}+R^{2}+\rho^{2}-2R\rho\cos{\phi}]^{3/2}}.
              \nonumber   \end{eqnarray}
  Here, $\textbf{r}=(\rho, \phi, z)$ are the coordinates of helium atoms. We position the coordinate origin on the resonator axis
at the center of the left end surface of the resonator.
 Using the DM (\ref{vp-1}) of helium atoms that arises due to the density gradient in a wave of
   second sound,  we obtain
         \begin{equation}
      \varphi_{b}(\textbf{R}) = \varphi_{0}\gamma_{b}(Z,R,\sigma)\cos(\omega_{2}t),
             \label{4-7}     \end{equation}
         \begin{equation}
      \varphi_0 \approx \frac{7\pi S_{7}\alpha_{T}d_{0}}{12\varepsilon \bar{R}^{2}} {\sss \triangle} T_{0},
        \label{4-8}     \end{equation}
   where $\sigma = R_{r}/L_{r}.$ We note that  only the quantity
       \begin{eqnarray}
       &&  \gamma_{b}(Z, R, \sigma) =  \sigma^{-1}\int\limits^{2\pi}_0 d\phi  \int\limits^{1}_{0} dz
   \int\limits^{1}_0 d\rho \rho \sin(\pi z) (Z-z) \times \nonumber \\
   &\times & [(z-Z)^{2}/\sigma^{2}+(R-\rho)^{2}+2R\rho(1-\cos{\phi})]^{-3/2}
    \label{4-9}     \end{eqnarray}
   depends on the observation point $Z, R$.
  In the formula (\ref{4-9}), $R$ and $\rho$ are normalized to $R_{r}$, and $z, Z$ --- to $L_{r}$.  In the new experiment, the resonator length $L_{r}=2.5\,cm,$  and its radius $R_{r}=0.5\,cm$. Therefore, $\sigma = 0.2$.
  We determined the dependence $\gamma_{b}$ on $Z$ at $\sigma = 0.2$ for several $R$ numerically by (\ref{4-9}) and present the results in Fig.~\ref{fig3}.
 For other $\sigma,$ the solution of (\ref{4-9}) can be found also numerically, and, approximately, the relation
$\gamma_{b}(Z,  R, \sigma)\simeq \gamma_{b}(Z, R, 0.2)\sigma/0.2$ holds for all $\sigma \leq 1$. The electrode occupies approximately a half of the end
surface of the resonator. So, the electrode is rather large, and we must average the potential over all points. But since the potential differs
insignificantly at points with different $R$ (see Fig.~\ref{fig3}), we can neglect this difference and consider the potential equal to that at the point
at the electrode center ($R=0, Z\rightarrow 0$).

 The field strength
 \begin{equation}
       \textbf{E}=-\textbf{i}_{z}\frac{\partial\varphi_{b}}{\partial Z}-\textbf{i}_{\rho}\frac{\partial\varphi_{b}}{\partial R}  =
        -\varphi_{0}\cos(\omega_{2}t)\left (\textbf{i}_{z}\frac{\partial\gamma_{b}}{\partial Z}+\textbf{i}_{\rho}\frac{\partial\gamma_{b}}{\partial R}  \right )
    \label{4-10}     \end{equation}
can be also determined numerically. With good accuracy, we have $\textbf{E}\approx \textbf{i}_{z}E_{z}$, $E_{z} =-P_{z}(\partial\gamma_{b}/\partial Z) \sin^{-1}(\pi Z/L_{r})$.
For example, under the experimental conditions with $T=1.8$~K and $\sigma$ = 0.2, we have
 $E_{z} \approx 0.1k_{B} {\sss \triangle} T_{0}\cos(\omega_{2}t)/|e|L_{r}\approx -2.5 P_{z}$ at the resonator center ($R=0, Z=0.5L_{r}$). At the beginning of the resonator axis
($R=0, Z\rightarrow 0$), $E_{z} \approx 0.06k_{B}{\sss \triangle} T_{0}\cos(\omega_{2}t)/|e|L_{r}\approx -1.4P_{z}/\sin{(\pi Z/L_{r})}\approx -1.4P_{z}L_{r}/\pi Z,$ i.e.,
the polarization is much less than the field strength, since $L_{r}/Z \gg 1$. In this case, it is necessary to take also into account the induced
polarization $\textbf{P} = \kappa\textbf{E}$ that determines the values of $\textbf{P}$ at $Z\rightarrow 0, L_{r}$. The numerical analysis indicates that
the consideration of the induced polarization leads to a slight change of
potential (\ref{4-7}): approximately by $1\%$. At the points near the resonator center, the change is greater, but the potential is low there.
Hence, such a correction can be neglected.
 We note that the consideration of the ``images'' of dipoles (arising in the electrode and walls of the resonator) in (\ref{4-9})
should change potential (\ref{4-7}) significantly, by several times (formula (\ref{4-9}) involves no images).

The proper DMs of polar molecules can be very different. As usual \cite{hip},
the two-atom molecules have $d_{p}\simeq 0.5|e| a_{B}$, though DMs
of some molecules are much less \cite{hip} (e.g., for molecules
$\textrm{CO}$ and $\textrm{NO}$, $d_{p}\simeq 0.05|e| a_{B}$). In field
(\ref{4-10}), the molecules are ordered and acquire a directed DM
(\ref{4-0}). At $d_{p}\simeq 0.5|e| a_{B}$ under the experimental
conditions with $T=1.8$~K, ${\sss \triangle} T_{0}\simeq 1\,mK,$ we
obtain that the directed DM is equal to
 $d_{E}\simeq 10^{-4}|e|a_{B}$, which exceeds a tidal DM $d_{0}$ (\ref{3-9}) by a factor of $\simeq 3$.
 In this case, the bulk polarization (\ref{vp-1}) of helium atoms at $T=1.8$~K is
$d_{z} \simeq   2.7\times 10^{-12}d_{0}$. This allows us to draw a significant conclusion that an admixture of polar molecules compensates
the bulk polarization of helium, if the share of an admixture (the ratio of the concentration of an admixture to the concentration of helium)
exceeds $\sim 10^{-12}$! It is a very small number: for example, the share of surface atoms of helium ($\sim 10^{-8}$)
is larger by 4 orders of magnitude. Most probably, helium contains always an admixture of polar molecules with a share of $\sim 10^{-12}.$
In this case, the bulk polarization is absent always in second-sound waves.

In Sec. V, one more kind of the bulk polarization of helium that is due to the dipole layer covering the
internal surface of a resonator is found. This polarization is less by one order of magnitude than that by (\ref{vp-1}), but the field strength for it by one order
of magnitude higher
than that by (\ref{4-10}). Therefore, such a polarization can be completely compensated by an admixture with a share of $\sim 10^{-14}$.
In what follows, this polarization will be omitted.

The case of the surface polarization related to polar molecules is more complicated. If a molecule is very close to the surface,
then it is attracted by the van der Waals forces, like nonpolar atoms. Moreover, since the molecule has proper DM, the mirror image of this DM appears in the
medium adjacent to helium (the image is exact if the medium is a metal and is weakened
if the medium is a dielectric), and we have an additional strong attraction of the dipole-dipole type. The orientation perpendicular to the surface is favorable
for a dipole. In this case, the interaction energy
\begin{equation}
    W=\frac{\textbf{d}_{1}\textbf{d}_{2}-3(\textbf{d}_{1}\textbf{n}_{12})(\textbf{d}_{2}\textbf{n}_{12})}{\varepsilon R_{12}^{3}}  = \frac{-2d_{1}^{2}}{\varepsilon R_{12}^{3}},
    \label{4-11}     \end{equation}
 where $\textbf{n}_{12}=\textbf{R}_{12}/R_{12}$,  $\textbf{R}_{12}$ is the radius-vector from the dipole $\textbf{d}_{1}$ to its
image $\textbf{d}_{2}.$ Here, we take into account that $\textbf{d}_{2}=\textbf{d}_{1}$ for a metal.
 Let $d_{1}=d_{p}\simeq 0.5|e| a_{B}$. If there is no solid layer of helium on the surface,
we have $R_{12}\simeq 2\bar{R}\approx 7.2\,\mbox{\AA}$ and $W\approx -60.5$~K. Such binding energy is much more
than the energy of any quasiparticle in helium, so that practically all atoms of an admixture
should be condensed on the surface. However, for molecules with a small DM $d_{p}\simeq 0.05e a_{B},$ we obtain $W\approx -0.6$~K. At such small
binding energy, quasiparticles will separate easily a molecule from the surface. A molecule can be held near the surface only by
the van der Waals forces, like atoms of helium themselves. Since the surface attracts atoms of helium weakly, we may expect that
it attracts atoms of the admixture in the same manner. Such an admixture should be floating freely in the bulk and should be able to damp the bulk
polarization of helium. However, due to a small $d_{p},$ the critical share of an admixture is greater by two orders of magnitude: $\sim 10^{-10}$.
If the surface attracts atoms of helium strongly, and one solid layer of atoms of helium is formed on the surface, then
$R_{12}\simeq 4\bar{R}\approx 14.4\,\mbox{\AA}.$ For $d_{p}\simeq 0.5e a_{B},$ we have in this case $W\approx -7.6$~K. Such binding energy
is easily overcome by a single roton. If the van der Waals interaction is taken into account, the value of $W$ increases by several times
(in modulus). But we can expect that the total $|W|$ is less than the energy of three-four rotons. In the case where
$T>1$~K and the number of rotons is great (and near the surface their concentration, apparently, is maximum \cite{dryF}),
the number of molecules of an admixture in the bulk will exceed that on the surface. The exact proportion can be calculated. Such admixtures
are also able to damp the bulk polarization. Some admixtures can form chemical bonds with a surface,
but we will not consider this case.

Since the dipoles of surface molecules of an admixture are oriented, on the average, perpendicularly to the surface, they create a double charged
layer and induce a potential on the electrode. We now estimate it in the case where an admixture covers all the surface
by a single atomic layer. The potential created by an admixture on the electrode is determined by formula
(\ref{5-7}), where the quantities $\bar{R}_{1||}^{-2}$ and $d_{1,z}$ must be replaced, respectively, by the surface concentration of an admixture
and by the $z$-component of DM of a molecule. In order to determine the variable signal from an
admixture, we need to know the derivative $\partial\varphi/\partial T$. By differentiating, we obtain
 \begin{equation}
            \frac{\partial\varphi}{\partial T} \simeq
        -\frac{2\varphi\alpha_{ad}}{3} + \frac{\varphi}{d_{1,z}} \frac{\partial d_{1,z}}{\partial T}.
      \label{4-12}    \end{equation}
Let us assume that the coefficient of thermal expansion
$\alpha_{ad}$ of an admixture is about $\alpha$ of liquid helium.
Then, at $d_{1,z}\simeq d_{p}\simeq 0.5|e| a_{B},$ only the first
term in (\ref{4-12}) is related to the appearance of an electric
signal that is by 4 orders of magnitude greater than the surface signal
(\ref{new5}) from atoms of helium. We do not know how the second
term in (\ref{4-12}) can be simply estimated. But we can expect
that the projection of a dipole reacts to a change of the
temperature stronger than the concentration, therefore the
contribution of this term should be significantly greater than
that from the first term in (\ref{4-12}). Thus, a single filled
layer of an admixture of polar atoms gives the surface signal up to $10$ mV, that
is greater by 4-5 orders of magnitude than the signal from atoms of helium. If atoms of an admixture fill the layer
only partially or, conversely, there are many layers, the signal
will change proportionally.

The estimates are very approximate for the surface polarization from polar atoms and are more exact for the bulk polarization. For the signal from the
admixture of $^3$He atoms, the estimates are quite reliable. In two last cases, the order of magnitude must be correct. It is seen from these estimates
that admixtures can strongly affect the signal, by increasing or decreasing it. As sufficiently unexpected, we mention the result on a possible
complete damping of the bulk polarization of helium by an admixture of polar molecules. Qualitatively, this damping is related to the extreme smallness
of the bulk polarization of helium: it is less by $\sim 12$ orders of magnitude than the surface polarization and by $\sim 16$ than the DM $d_{p}$ of polar molecules.
Therefore, the latter can easily compensate the bulk polarization. To clarify the role of admixtures, a more detailed analysis should be performed.
The principal question is whether the critical concentration of bulk polar admixtures is reached.

We note that the experimental signal can be caused, in principle,
by a layer of polar molecules on the electrode  with the
concentration  $\sim 10^{-4}$ of that of helium. This is seen
from the above estimates. But then the signals must be different
in different experiments, which was not observed. Therefore, the
signal is related namely to atoms of helium, and the theory of
such a signal (see above) is in good agreement with experiments.

It would be of interest to experimentally investigate the role of admixtures.

        \section{The electric signal induced by a first-sound wave}

         According to the model, the signal in a wave of second sound arises for two reasons.
         Due to the isotropy breaking in the system, the helium
         atoms are strongly polarized in the region near the electrode, and the polarization oscillates due to
           oscillations of the density of helium. In fact, the signal is determined by the low-intensity first sound accompanying
           second sound. Then the electric signal undoubtedly should be
           observed after excitation of a standing half-wave of the first (as the main) sound in
           helium. However, such a signal was not observed in the experiment \cite{rub}.
         Below, we will calculate the expected magnitude of the signal.

          In a  standing half-wave of first sound in the harmonic mode, the pressure varies according to the law
           \begin{equation}
       p = p_0 - 0.5 {\sss \triangle}  p(z) \cos(\omega_{1} t), \
         {\sss \triangle}  p(z) =  {\sss \triangle}  p_{0} \cos(z\pi/L_{r})
     \label{1s-1}     \end{equation}
      (and similarly for the density);  $ {\sss \triangle}  p(z)$ and
      $ {\sss \triangle}  \rho(z)\equiv  {\sss \triangle}  \rho_{0} \cos(z\pi/L_{r})$ are the amplitudes of oscillations
      of the pressure and density of helium, respectively, at points with coordinate $z$.

          The signal observed in a wave of second sound can be written in the form (\ref{new5})
          just from the dimensional analysis. The experiment gives $a \approx -1$.
      The quantity $a$ must be calculated in theory. Starting from the dimensional consideration,
      we have, for first sound,
         \begin{equation}
     \frac{ {\sss \triangle}  \varphi(z)}{ {\sss \triangle}  p(z)} = \frac{b}{|e|n}
      \approx b \cdot 2.89\cdot 10^{-5}\frac{\mbox{V}}{{\rm atm}},
     \label{1s-2}     \end{equation}
    where $n$ is the concentration of He~II\@.
   Since the near-surface polarization arises on atomic scales,
       we may expect that $|b| \sim 1$.

        Similarly to second sound, first sound induces surface and bulk signals.
         We will determine the surface signal analogously to the case of second sound,
         by expanding the potential in the density (instead of temperature)
         and assuming that
         $ {\sss \triangle}  p$ is the same in different near-surface layers. In this way, we obtain:
                   \begin{equation}
         b_{2D} = b_{1}+b_{2}+b_{3}+\ldots,
                \label{1s-3}     \end{equation}
        \begin{eqnarray}
   b_{1} &\approx &  \frac{|e|\varphi_{1}\rho_{0}}{3m_{4}\rho_{1}u_{1}^{2}(p_{1})}
   \left \{ 2   + 7\frac{d_{\rm h}}{d_{1}}\left (\frac{1}{1+\bar{R}_{2}/\bar{R}_{1} } + \right. \right. \nonumber \\
    &&{} + \left. \left.  \frac{\rho_{1}u_{1}^{2}(p_{1})}{\rho_{2}u_{1}^{2}(p_{2})}\frac{1}{1+\bar{R}_{1}/\bar{R}_{2} }
      \right )\right \},
       \label{1s-4}     \end{eqnarray}
        \begin{eqnarray}
   b_{j\geq 2} &\approx &  \frac{|e|\varphi_{j}}{3m_{4}u_{1}^{2}(p_{j})}
   \left \{ \frac{2\rho_{0}}{\rho_{j}}   + \right. \label{1s-5}  \\
  &&{} + \left. 7\frac{\frac{u_{1}^{2}(p_{j})}{u_{1}^{2}(p_{j-1})}\left (\frac{\rho_{j-1}}{\rho_{0}} \right )^{4/3}
   - \frac{u_{1}^{2}(p_{j})}{u_{1}^{2}(p_{j+1})}\left (\frac{\rho_{j+1}}{\rho_{0}} \right )^{4/3} }{
      \left (\frac{\rho_{j-1}}{\rho_{0}} \right )^{7/3} - \left (\frac{\rho_{j+1}}{\rho_{0}} \right )^{7/3}}
        \right \},
          \nonumber  \end{eqnarray}
            where $p_{0}$ and $\rho_{0}$ are the bulk values. At the first solid layer, $p_{2}=13$~atm and
       $p_{j\geq 3}={\rm svp}$, we obtain $b_{1}\approx 0.54$ for Au and $b_{1}\approx 0.58$ for Cu.
     For both metals, $b_{2}\approx 0.71$, $b_{3}\approx 0.3$, and $b_{j\geq 4}$ are small and form $b_{3D}$
     (the bulk signal).
     This yields $b_{2D}\approx 1.55$ for Au and
        $b_{2D}\approx 1.6$ for Cu. As $p_{2}$ increases from 0 to $20$~atm, the value of $b_{2D}$
        increases weakly, approximately by 10\%. We note that, due to high pressure
        in the first layers, $ {\sss \triangle}  p$ in them can be smaller than the bulk value,
        but this circumstance does not change the order of $b_{2D}$, according to estimates.

     We can estimate the bulk signal at the electrode with the coordinate $z=0$ in a wave of second sound
     by using equations (38) and (40) from Ref.~\onlinecite{volpol} (another bulk signal (\ref{vp-6}), from the DL on the resonator surface,  is weaker by one order of magnitude  and is omitted). Using        $ \alpha{\sss \triangle}  T  \approx -  {\sss \triangle}  \rho/\rho$, we obtain
     \begin{equation}
      b_{3D} \approx
     \frac{7\pi S_{7}\gamma(R_{r}/L_{r})|e|d_{0}}{6\varepsilon\bar{R}_{0}^{2}u_{1}^{2}(p_{0})m_{4}}
       \approx 0.48\gamma(R_{r}/L_{r}).
        \label{1s-6}     \end{equation}
              By introducing the potential $\varphi_{{\rm h}0} = -4\pi d_{\rm h}(\rho_{0})/(\varepsilon \bar{R}_{0}^{2})$,
         we can represent $b_{3D}$ in the form $b_{3D} \approx
     \gamma(R_{r}/L_{r})\times 7|e|\varphi_{{\rm h}0}\left[6u_{1}^{2}(p_{0})m_{4}\right]^{-1}$
     which is similar to the expressions for $b_{2D}$.
     The factor $\gamma_{b}(R_{r}/L_{r})$  describes the cutting of a bulk signal. The value of such $\gamma$ is
     determined in Ref.~\onlinecite{lt1}: for short and long resonators from the
        experiment \cite{rub}, one has $\gamma\approx 1.4$ and $\gamma\approx 0.05$, respectively. Then
        $b_{3D}\approx 0.66$ and $b_{3D}\approx 0.024$, respectively. The total values are
        $b=b_{2D}+b_{3D}\approx 2.2$ and $1.57$,  respectively, for Au,
         and $b\approx 2.26$ and $1.62$,  respectively, for Cu.
       The sign of $b$ determines the signal polarity.

     First sound was studied \cite{rub} for a short resonator;
     therefore, the signal at the electrode with $z=0$ can be described by (\ref{1s-2})
     at $b\approx 2.2$. The problem consists in the estimation of $ {\sss \triangle}  p$.
      In experiments, $ {\sss \triangle}  p$ was not measured. We know \cite{r3} only the maximum power of an
        acoustic emitter ($w_{1 \rm em}\approx 5$~mW) and  the $Q$-factor for this power ($Q_{1}\simeq 40$).

        Firstly, let us consider the experiment with second sound for which the value $ {\sss \triangle}  T_{0}$ was measured
        and  let us evaluate this quantity theoretically.
        This can be done in the usual way, by equating the energy of a temperature wave
         \begin{equation}
       {\sss \triangle}  E \approx  \frac{\partial E}{\partial T}\langle  {\sss \triangle}  T \rangle +
      \frac{\partial E}{\partial p}\langle  {\sss \triangle}  p \rangle
                    \label{1s-7}     \end{equation}
        to the pumping energy multiplied by the $Q$-factor: $\zeta w_{\rm em}\tau Q$.
          Here, the average is over time and $z$, the quantity $\zeta$ is the coefficient of attenuation
         (the ratio of the emitted power to the total power of a heat emitter), and $\tau$ is the wave period.
         It is worth noting that, while considering the energy balance for
          a standing wave of first or second sound,
            $T$ and $ p$ should be determined from the minimum values in the wave. The energy in a resonator only flows
       from one side to another side. Its value is not changed, and the losses are exactly compensated by the pumping.
        Just this energy can be equated to  $\zeta w_{\rm em}\tau Q$. But if $T$ and $ p$ are taken
        from the averaged values,
        as it is usually done \cite{pat,xal},
        then the main part of the energy (linear approximation) is lost, because it is nullified. Therefore, we consider
        below $ {\sss \triangle}  T(t,z) = 0.5 {\sss \triangle}  T_0 \left[ 1 - \cos(\omega_{j} t)\cos(z\pi/L_{r}) \right]$
         (and similarly for $ p$), where $j$ is the sound number.

           Using the equations \cite{pat,xal}
            \begin{equation}
     E=E_{0}+\textbf{v}_{s}\textbf{j}_{0}+\rho\textbf{v}_{s}^{2}/2, \quad \textbf{j}_{0}=\rho_{n}(\textbf{v}_{n}-\textbf{v}_{s}),
              \label{1s-8}     \end{equation}
      \begin{equation}
      dE_{0}=TdS+\mu d\rho + (\textbf{v}_{n}-\textbf{v}_{s})d\textbf{j}_{0}
              \label{1s-9}     \end{equation}
        ($E$,  $S,$ and $C$ are given per unit volume) and the thermodynamic relations,
         we obtain
         \begin{equation}
       {\sss \triangle}  E \approx     (C_{p}-\mu\rho\alpha)\langle  {\sss \triangle}  T \rangle
        + \left (\frac{\mu}{c_{1}^{2}}-T\alpha + \frac{TS}{\rho c_{1}^{2}}\right )\langle  {\sss \triangle}  p \rangle.
              \label{1s-10}     \end{equation}
       The chemical potential $\mu$ can be determined from the equations
        \begin{eqnarray}
      \mu &=& \left. \frac{\partial F_{0}}{\partial \rho}\right|_{T} \approx
       c_{1}^{2} \left. \frac{\partial F_{0}}{\partial p}\right|_{T} = \nonumber \\
        &=&  c_{1}^{2} \left. \frac{\partial }{\partial p}\right|_{T}\left[ E_{0}(T=0)-k_{B}TN_{\rm r}
         - \frac{\pi^{4}}{108}k_{B}TN_{\rm ph}\right] = \nonumber \\
         &=&           \mu(T=0)+ {\sss \triangle}  \mu (T),
              \label{1s-11}     \end{eqnarray}
                  \begin{equation}
       \mu(T=0)=c_{1}^{2} \left. \frac{\partial E_{0}(T=0)}{\partial p}\right|_{T} \approx \frac{\epsilon_{0}}{m_{4}} \approx -0.27c_{1}^{2},
              \label{1s-12}     \end{equation}
        \begin{eqnarray}
        {\sss \triangle}  \mu (T) &\approx & -c_{1}^{2}\cdot 10^{-3}\left (1.7\left (\frac{T}{1.6\,K}\right )^4 + \right. \label{1s-13} \\
      &+& \left.  2.9\sqrt{\frac{T}{1.6\,K}}
              \exp{\left [-\frac{\Delta(T)}{T}+\frac{\Delta(1.6\,K)}{1.6\,K}\right ]}\right ).
                \nonumber   \end{eqnarray}
              Here  $\Delta(1.6\,K)\approx 8.42\,$K, and  $\epsilon_{0}=-7.16\,$K is the ground-state energy per atom.
              In Eq. (\ref{1s-13}) we used the data \cite{es,and2} on the dependence of the parameters of phonons and rotons
              on $p$.

         For second sound, $ {\sss \triangle}  \rho \approx - \alpha\rho{\sss \triangle}  T,$ and $ {\sss \triangle}  p$ is small.
          Equating energy
        (\ref{1s-10}) to the pumping one, we get
          \begin{equation}
      (C_{p}-2\mu\rho\alpha) {\sss \triangle}  T_{0}/2 \approx \zeta_{2} w_{2\rm em}\tau_{2} Q_{2}/\Omega,
              \label{1s-14}     \end{equation}
         where $\tau_{2} = 2L_{r}/c_{2}$,     and the factor $1/2$ appears on the left-hand
         side due to the averaging:  $\langle {\sss \triangle}  T(t,z)\rangle  = \frac12 {\sss \triangle}  T_0$.
         The linear dependence between $w_{2\rm em}$ and $ {\sss \triangle}  T_{0}$ was observed
         in experiments up to the critical value of the heat flow
          $w_{2\rm em, c}Q_{2}/S_{h} = 4~\mbox{W}\,\mbox{cm}^{-2}$ at $Q_{2}\simeq 2000$;
          here $S_{h}$ is the area of the heater.
           Whence we obtain $ {\sss \triangle}  T_{0} \simeq 0.037\zeta_{2}$~K for
         a short resonator (with the volume $\Omega =L_{r} S_{r} \approx  0.82\,\mbox{mm}^3$)
         at $T=1.6$\,K ($S_{r}$ is the area of the resonator,
         $S_{r} \approx S_{h}$).
          The value of $\zeta_{2}$ is unknown.
          A wave of second sound includes a small admixture of first sound
          [in (\ref{1s-10}), the acoustic wave energy $\sim 4\%$ at $T=1.6$~K]\@.
        Both sounds are coupled by the relation $ {\sss \triangle}  \rho \approx - {\sss \triangle}  T \alpha\rho $.
        The pumping is only a heat one; therefore $\zeta_{2}$ is determined by the
         heat inertia \cite{pesh2}
         caused by the finiteness of both the heat passage duration through a heater and the durations of the
         creation and the diffusion of a roton and a phonon (a roton needs a time to depart from the wall and to liberate
         the place for the next one). In addition, it should be taken into account that, in the presence of a heat pumping, the
          Kapitsa jump of $T$ arises in the near-surface
         layer of He~II (with the thickness $\lesssim 10^{-3}\,{\rm cm}$) near a heater \cite{kap,xal-skacokT},
            \begin{equation}
       {\sss \triangle}  T_{\rm K} \approx \frac{Aw}{T^{3}S_{h}},
                    \label{1s-K}     \end{equation}
     where $w$ is the power  of the heater. For copper, $A \approx 5$$-$$50~\mbox{K}^{4} {\rm cm}^{2}/{\rm W}$
     (see Ref.~\onlinecite{enz}),
      and the values of $A$ are close for the other metals. Therefore, at $T= 1.4$~K and
      $w_{2\rm em}/S_{h} \lesssim 2\,\mbox{mW}/\mbox{cm}^2$, we have $ {\sss \triangle}  T_{K} \approx 3.6$$-$$36$~mK\@.
      Near $T_{\lambda}$, the law (\ref{1s-K}) is violated.
      In waves of second sound, we observe two types of oscillations of $T$:
      bulk ones and oscillations in the near-surface layer [due to the
       jump (\ref{1s-K})]. At the resonance, these oscillations
      must be consistent. Since $ {\sss \triangle}  T_{\rm K}$ sets the amplitude of variations of $T$, the relation
      $ {\sss \triangle}  T_{0}\lesssim  {\sss \triangle}  T_{\rm K}$ should be valid.
      At the resonance, we may expect that $ {\sss \triangle}  T_{0}\sim  {\sss \triangle}  T_{\rm K} $
      (in the ideal system, where the heater, walls, and the thermometer are made of the same material,
      it is probable that $ {\sss \triangle}  T_{0} = {\sss \triangle}  T_{\rm K} $)\@.
      From the experiment, we have $ {\sss \triangle}  T_{0} \simeq 3$~mK at $T=1.4$~K, and
               $ {\sss \triangle}  T_{0} \simeq 1.3$~mK at $T=1.6$~K\@. One can see that, at $T=1.4$~K,
               the quantity    $ {\sss \triangle}  T_{0}$ is of the same order as  $ {\sss \triangle}  T_{\rm K} $,
               being several times smaller.  More exact data of
            a new experiment \cite{r3} indicate that
             their dependencies on the temperature are also close. The experimental value of
               $ {\sss \triangle}  T_{0}$ at $T=1.6$~K
       is obtained at $\zeta_{2} \simeq 1/29$. It is apparent that $\zeta_{2}$ is determined by the value
       of $ {\sss \triangle}  T_{\rm K}$, i.e., by the Kapitsa jump.

                  We note that formula (\ref{1s-14}) without both $\zeta_{2}$ and a correction with
          $\mu$ was obtained earlier in Ref.~\onlinecite{pesh1} in a different way. In Ref.~\onlinecite{pesh2},
          it was observed that the attenuation is almost absent
          ($\zeta_{2}\simeq 1$) for second sound at frequencies
          $\nu \sim 100$$-$$500$~Hz. However, strong attenuation occurs at $\nu \gtrsim 5$~kHz,
which corresponds to the result obtained above:  $\zeta_{2} \approx 1/29$ for $\nu
\approx 10$~kHz.

       Let us consider first sound.
       From the system of equations for first and second sounds (see Ref.~\onlinecite{pat}, Chap.~1, $\S 7$),
       it is easy to obtain       that
                  \begin{equation}
       {\sss \triangle}  T \approx -  {\sss \triangle}  p\cdot \alpha T/C_{p}.
                    \label{1s-15}     \end{equation}
      Using Eqs.~(\ref{1s-10})--(\ref{1s-13}) and (\ref{1s-15}) and
             equating $ {\sss \triangle}  E$ to the pumping $\zeta_{1}w_{1\rm em}\tau_{1} Q_{1} $
             with $\tau_{1} = 2L_{r}/c_{1}$, we get
         \begin{equation}
      \frac{ {\sss \triangle}  p_{0}}{2}\left (\frac{\mu}{c_{1}^{2}}-2T\alpha + \frac{TS}{\rho c_{1}^{2}}
      + \frac{\mu\rho T\alpha^2}{C_{p}}  \right ) \approx   p_{1E},
              \label{1s-16}     \end{equation}
      where $p_{1E}=\zeta_{1}w_{1\rm em}\tau_{1} Q_{1}/\Omega$.
           The quantity $\zeta_{1}$ and the critical flow $w_{1\rm em, c}Q_{1}/S_{h} $ are unknown, and the analysis performed for second sound indicates that they involve a sufficiently complicated physics.
       We only note that an acoustic emitter was the membrane of a headphone, in which the electric signal $w_{1\rm em}$
       is transformed
      into an acoustic one only partially. In addition, the frequency of first sound is larger
      by a factor of 11.5 than that of second sound,
       which favors a decrease of $\zeta_{1}$. For estimates, let us take
       $\zeta_{1}=\zeta_{2}=1/29$ and $w_{1\rm em, c}Q_{1}/S_{h} = w_{2\rm em, c}Q_{2}/S_{       h} = 4\,\mbox{W}\,\mbox{cm}^{-2}$. Then  $p_{1E}\approx  \zeta_{1}\cdot 3.4 \cdot 10^{-3}\,$atm and
          $ {\sss \triangle}  p_{0} \approx -0.026\zeta_{1}\,{\rm atm} \approx -9\cdot 10^{-4}\,{\rm atm}$ (the minus sign indicates that the internal energy decreases in the region of enhanced
       pressures, and we will use $| {\sss \triangle}  p_{0}|$ in what follows).
       In this case, the pressure of saturated vapor is
        $p_{0}\sim 2.5\cdot (10^{-3}\mbox{$-$}10^{-2})\,$atm at $T=1.4\mbox{$-$}1.8$~K.
         With regard for (\ref{1s-2}) at $b\approx 2.2$, we obtain
       a signal $ {\sss \triangle}  \varphi(z=0) \approx 57\,$nV, which is about the minimally registered signal $10\,$nV
        by the order of magnitude. Apparently, a significantly smaller $ {\sss \triangle}  p_{0}$,
        less than $ 10^{-4}\,$atm, was attained in experiments,
        and $ {\sss \triangle}  \varphi$ turned out to be below the threshold of registration.

      It is interesting that $\sim 97\%$ of the energy of acoustic oscillations
      [the second term in (\ref{1s-10})] are given by the term
      $\mu(T=0)\langle  {\sss \triangle}  p \rangle /c_{1}^{2} \approx -0.3\langle  {\sss \triangle}  p \rangle$ which
       is the energy of pulsations of the ground state of He~II\@.
       These pulsations are running together with the movement of quasiparticles,
        but the energy related to quasiparticles is smaller by two orders of magnitude.
        Thus, the negative energy of the ``vacuum'' dominates.

         We note that, according to (\ref{1s-10}),
        a standing half-wave of first sound is always accompanied by a low-intensity wave of second sound (heat oscillations) bearing $\sim 1.5$\% of the energy.
       However, heat oscillations are easily absorbed by the walls, so that the heat losses are significantly greater than $1.5\%$.
       In this case, the heat pumping is absent, and the transformation of the acoustic energy to quasiparticles is slow.
        Therefore, a wave of second sound gives possibly the main contribution to the attenuation of first sound. If this is true, then
         $ {\sss \triangle}  p_{0}$ can be increased by two measures: by completely suppressing a heat
       wave (to the level of fluctuations) or, on the contrary, by its pumping. In this case, the $Q$-factor
       must sharply increase.
        The first measure can be realized by approaching the temperature at which
        $\alpha \approx 0$ ($T\approx 1.12$~K  or $1.18$~K, according to  Refs.~\onlinecite{es}
       and \onlinecite{pat}, respectively). In the second case, it is necessary
         to induce a heat pumping (with the frequency and phase of first sound) in addition to the acoustic one.
         One more measure consists in the maximum decrease of the frequency with increase in the resonator length.
         Finally, it is possible to use the method of filtration for the generation of first sound \cite{pesh1,pesh2}.
         According to Ref.~\onlinecite{pesh2}, this method allows one to attain
         $ {\sss \triangle}  T_{1}\simeq  {\sss \triangle}  T_{2}/6$
         and to obtain large $ {\sss \triangle}  p_{0}$ up to $0.1$~atm.
         In these case, the signal $ {\sss \triangle}  \varphi$ {\em should be strong\/} (up to $ 1000\,$nV) {\em and  observable\/}.

       It is worth noting that the absence of any information about
       $\zeta_{1}$ and $w_{1\rm em, c}Q_{1}/S_{h}$ requires to measure the value of $ {\sss \triangle}  p_{0}$
       directly or to measure $ {\sss \triangle}  T_{0}$ and then to determine $ {\sss \triangle}  p_{0}$
         from (\ref{1s-15}).

         We believe  that the electric signal arises in a wave of second sound
         because such a wave is always accompanied by
          first sound. In this case, second sound is only a
          way to generate first sound, and, under certain conditions,
          this way is more efficient than the direct acoustic generation of first sound.
          Undoubtedly, there exist methods
           of direct excitation of first sound which allow one to obtain high-amplitude
           pressure oscillations and a large signal $ {\sss \triangle}  \varphi$.
             Due to the coupling of the two sounds, the nature
           of the electric signal can be understood only
           by simultaneously studying both sounds. Therefore, the further studies
           of first sound appear to be of great importance.

                        \section{Discussion of the results and a new experiment}

        We note some specific features of the electric signals for first and second sounds.
         According to the model, the signal
           has the same nature for both sounds and is related to oscillations of the density.
           The difference consists in the following: for first sound, oscillations of the density are induced directly,
           whereas for second sound they are generated indirectly  due to the weak coupling of the two sounds.
             For both sounds, the
         signal consists of the surface and bulk parts.
          The first part dominates and does not depend on the size of the resonator and on the temperature.
          Moreover, the signal for first sound does not depend on $T$ irrespective of the consideration of the
          exhaustion of $\rho_{s}$ at the wall, whereas it is necessary to take the behavior of $\rho_{s}$ into account for second sound.
           It is also clear that the signal for first sound is not related to
           superfluidity and must be observed above $T_{\lambda}$.
           For second sound, some limitation consists in that the connection between $\nabla T$ and $\nabla\rho$ is unclear above $T_{\lambda}$.
           Moreover, the quasiparticles are poorly defined above $T_{\lambda}$, so that second sound
           cannot be excited. Nevertheless, the effect should be observable above $T_{\lambda}$
           if stable oscillations of the temperature can be induced and if $\nabla T$ generates $\nabla\rho$.

           It is interesting that the constants $a$ and $b$ are close to unity
           in the dimensional formulas for both effects.
           This is apparently related to the dominance of the surface polarization  induced
           in several atomic layers:
            the constants $a$ and $b$ must be of the order of unity if the effect arises on atomic scales.

                       In addition, the theory predicts a number of peculiarities. Namely,
              for both sounds, the signal $ {\sss \triangle}  U$  must significantly differ for electrodes belonging to
         different groups: those forming two solid layers of helium, those forming one
         solid layer, those that do not form solid layer, but are wetted,  and those that are nonwetted.
         In this case, the signals must be close for electrodes made of different metals of the same group,
         although the binding energies of the $^4$He atom are significantly different for
            different metals of the group. Such a closeness of signals is mainly related
            to the fact that the metal-induced polarization of a $^4$He atom located near the electrode
             gives a significantly smaller contribution to the signal than the polarization which is induced by the half-space with helium
             and does not depend on the kind of a metal.
          This can be  verified in experiments. The three electrodes from the experiment  \cite{rub}
          probably belonged  to the same group.
            Therefore, the signals were practically identical for them.
          Additional differences can be for metals with the
      hexagonal lattice
      and for intermetals with a number of unusual properties,
      including a large coefficient $\alpha$ of thermal expansion
      approaching that for He~II\@.

            The coefficients $a$ and $b$ are calculated with an error of about one order of magnitude.
            It is caused by the error of the
            correlation function $g(r)$ at small $r$ and by the neglect of
            anisotropy of $g(\textbf{r})$ near the surface, as well as by the neglect of the
            difference between the longitudinal
            and transverse coefficients of linear expansion for the first layers.
            This leads to the error of $a_{j}$ and $b_{j}$
            up to several times.
             A considerable error for $a=a_{1}+a_{2}+a_{3}$ is introduced by the difference of signs of $a_{1}$ and $a_{3}$ which are
              almost identical in absolute value. The summary error of $a$ is about one order of magnitude.
               For $b$, all components $b_{j}$ are of the same sign and order.
               The quantities $a_{j}$ are also sensitive to both the value of $p_{2}$ and the distribution of $T$
               in the first layers, whereas $b_{j}$ are insensitive to them. Therefore, $b$ can be calculated more reliably than $a$ with an error up to several times.

                  In calculations, we did not consider the microroughness of
           the metal surface and the oxide film. The microroughness should
           not be significant since the value of $d_{\rm  mir}$ must be the same for different charge distributions in the metal
           and  equal to $d_{\rm  mir}$ from the exact mirror image. But an oxide film
             can significantly change the results, and this case must be examined specially.  Some
           electrodes are easily oxidized in air, and
           their surface should be specially prepared to prevent the formation of an
           oxide film.

            We also did not take into account  the  inertial polarization of $^4$He atoms arising
         due to their collisions with the metal surface.

                    We were interested only in metals with a cubic lattice.
           However, a lot of metals has a hexagonal lattice, for which the Wigner--Seitz cell has a quadrupole
            moment creating an electric field. Under the action of this field, the
            conduction electrons are redistributed in a metal so that the field inside and outside the metal
            become zero due to the
             formation of a dipole layer on the metal surface.
            This dipole layer will additionally contribute to $d_{1}$ (the DM of $^4$He atom at
           the metal surface). The induced DM is very large, $\sim 1000 d_{\rm  mir}$, but it is compensated
           by DMs from the quadrupoles of cells.
           However, due to the exchange interaction, the compensation is not complete, which can considerably change $d_{1}$ and $a$.

            Let us consider the question about a jump of $T$ of helium in several near-surface layers. The existence of such a jump follows
           from the reasoning of Sec. III and allows us to explain the $T$-independence of the
           signal $U$ in the experiment with second sound. As was mentioned above, a jump of $T$ in near-surface layers of helium
           was discovered by  Kapitsa \cite{kap} and is related to
            the high heat conductivity of helium.  But the mechanism considered in Sec. III predicts a jump of $T$
            in a thinner layer $\sim 10 \,\mbox{\AA}$ at the wall. This jump has a different nature and is related to the condensation of rotons on the wall.
                This mechanism seems to be important, because it concerns also the other properties such as the exhaustion
             of $\rho_{s}$, dry friction, and $T$-independence of $U$. This question deserves a separate
             more strict study. The key assumption of our consideration is that the condition
               $\rho_{s}=0$ on the wall yields the equality $T=T_{\lambda}$ on the wall (see Sec. III). This allows us to conclude that,
          in the first layers at the wall, $T$ must smoothly vary from the bulk temperature far from the wall to
         a higher $T$ of the wall equal to $T_{\lambda}$ at the pressure on the wall.
             The other reasonings lead to the same conclusion. Rotons at the wall have energy that is less by
       2\,K than those of bulk rotons. Hence, their concentration is higher by several times. However, these 2D-rotons
        involve atoms of several layers nearest to the wall into the motion. Hence, the temperature of the medium at the distance of 2-3
        atomic layers from the wall is determined not only by bulk quasiparticles, but by surface ones as well.
         But the number of 2D-rotons is greater by several times
        than that of bulk ones. Therefore, 2D-rotons significantly  increase the temperature of the first layers.
        In addition, we should expect the presence of the exchange by quasiparticles between the wall and the bulk.
        Since the energy and the concentration of 2D-rotons and phonons differ from their values for bulk quasiparticles, 3D-rotons and phonons near the wall undergo
         the different actions from the sides of helium (3D-rotons and phonons) and the wall
         (2D-rotons, 2D-phonons, and atoms of the wall). In other words, the equilibrium concentration of
        3D-quasiparticles at the wall must differ from the concentration far from the wall. Hence, we have $\nabla_{z}T\neq 0$ near the wall.
        The thickness of the layer with $\nabla_{z}T\neq 0$ is $\sim$ the effective size of a roton
         ($ \sim 3$ atomic layers).
          This is in agreement with the experimental data on the third-sound \cite{3sound1,nul1}
       which give the thickness of the layer with $\rho_{s} \approx 0$ equal to $2\div 4\,\mbox{at.
       layers}$ at $T=1.4\div 1.8\,$K.
                  From the viewpoint based on the symmetry,
         the gradient of $T$, like the gradients of pressure and polarization, is related to the system isotropy breaking at the wall.

                   At the present time, a \textit{new experiment} is carried on with second sound
         \cite{r3} to determine the dependence of the
          electric signal on the coordinate ($Z$) along a resonator and
          the temperature.
           We will determine the possible results of the experiment within the present model.

\begin{figure}[ht]
\centerline{\includegraphics[width=85mm]{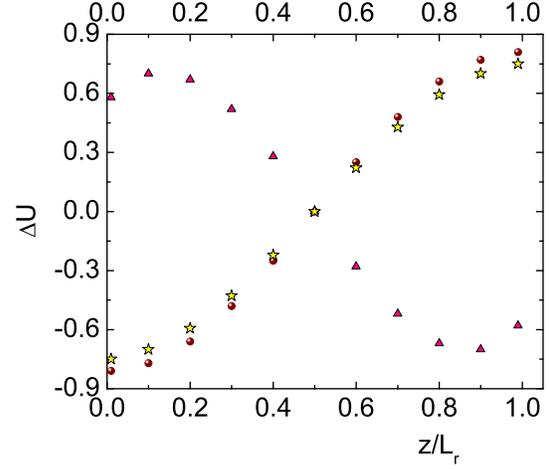}
}
\caption{Theoretical potential difference ${\sss \triangle}  U(z)$ (normalized to $k_{B}/2|e|$) between a ring electrode
with coordinate $z$ and the ground. The symbols $\bullet\bullet\bullet$ show the $z$-dependence for the surface signal
(\ref{6-1}), and $\triangle\triangle\triangle$ are related to the bulk signal (\ref{6-4}) increased by 10 times for clearness.
$\star\star\star$ is the summary curve of the bulk and surface signals. The bulk signal is presented under experimental conditions with
$\sigma = 0.2$ and $T=1.8$~K, and the damping due to images and admixtures is not taken into account in this signal.
 \label{fig4}}
\end{figure}

            Since the model considers the signal as mainly the surface one,
            the signal dependence on
           $z$ must be completely determined by the dependence (\ref{5-8}) of
            the temperature of He~II on $z$. Above, we have determined the potential $\varphi$ at
            the end electrode ($z=0$). In
           the experiment, the potential difference $U$ between two
           electrodes is measured.  The amplitude
           $ {\sss \triangle}  U$ of the potential difference between the
           electrode with coordinate $z$ and the ground
           is given by formula  (\ref{new5}), where $a$ depends on $T_{0}$ and surface properties (and is independent of
            $z$), and $ {\sss \triangle}  T$ depends on $z$ by (\ref{5-8})
             as $ {\sss \triangle}  T(z)= {\sss \triangle}  T_{0}\cos(\pi z/L_{r})$. Whence
           \begin{equation}
         {\sss \triangle}  U(z) = a\frac{k_{B}}{2|e|} {\sss \triangle} T(z) =
        a\frac{k_{B}}{2|e|} {\sss \triangle} T_{0}\cos(\pi z/L_{r}).
    \label{6-1}     \end{equation}
        The potential difference between identical ungrounded
    electrodes with the coordinates $z$ and $0$ is equal to
   \begin{equation}
      {\sss \triangle}  U(z) = a\frac{k_{B}}{2|e|} {\sss \triangle} T_{0} \left[ 1-\cos(\pi z/L_{r})\right].
    \label{6-3}     \end{equation}
    Two last formulas are true for the surface signal. To make
comparison with the experiment, we need to know also the
$Z$-dependence of the bulk signal that is registered by a ring
electrode positioned on the internal surface of the resonator. The
amplitude of the potential difference between this electrode with
coordinate $Z$ and the ground follows from formulas
(\ref{4-7})--(\ref{4-9}) after the replacement $R\rightarrow
R_{r}, \cos(\omega_{2}t) \rightarrow 2$:
         \begin{equation}
  {\sss \triangle} U_{b}(z) =   \varphi_{b}(Z) = 2\varphi_{0}(T)\gamma_{b}(Z,R_{r},\sigma).
               \label{6-4}  \end{equation}
The dependence of $\gamma_{b}$ on $Z$ at $R=R_{r}$ for the experimental value $\sigma = 0.2$ is given in Fig.~\ref{fig3}.
In Fig.~\ref{fig4}, we present the $Z$-dependencies of the surface (\ref{6-1}), bulk (\ref{6-4}), and
total signals. For the surface signal, we took $a \approx -0.8$ for copper (see Sec. IV).

As seen from Fig.~\ref{fig4}, the dependencies for the surface and bulk
signals are somewhat similar, but are different. Their similarity
is due to the fact that the main contribution to the bulk
potential is given by the region of helium near the electrode. The
difference is related to that this region has a macroscopic
thickness ($\sim 2R_{r}$), whereas the thickness corresponding to the
surface signal is equal to several atomic layers.

At the derivation of the potential difference (\ref{6-4}), we did
not take into account that the electrodes in the new experiment
are covered by a dielectric film. In addition, it is shown in
Sec. VI that the bulk signal is completely damped by an
admixture of polar molecules, if its partial concentration $\gsim
10^{-12}$. Such an admixture is always present, most likely, in
helium, which nullifies the bulk signal. In Fig.~\ref{fig4}, we did not
consider the bulk signal (\ref{vp-6}) arising due to DL on the
electrode surface. Such a signal is analogous to the surface
signal by its properties and is less by two order of magnitude by intensity.
Therefore, we omit it. In addition, this bulk signal is easily
damped by polar admixtures at their concentrations $\gsim
10^{-14}$. Moreover, formula (\ref{6-4}) does not involve the
images of dipoles in a metal. On the resonator walls parallel to the
resonator axis, these images-dipoles are directed oppositely to
dipoles in helium and, therefore, decrease significantly the
potential on ring electrodes (probably by several times). For the
electrode on the plane end surface, the images-dipoles are
directed as the dipoles in helium. Therefore, they increase the
signal up to two times.

The curves in Fig.~\ref{fig4} are the \textit{prediction} of the model. The same dependencies on $Z$ are obviously valid for the surface
signal from first sound.    The same dependencies on $z$ are obviously valid for the surface signal from first sound.

       Since the electrode is covered by a dielectric, the potential on the electrode surface
      is induced by three sources: the polarized
     first layers of helium on the dielectric surface and layers of a dielectric on the surfaces of helium and a metal.
     We will determine the electrode potential analogously to (\ref{5-7}). With regard for only the first layers, we obtain
     \begin{equation}
  \varphi \approx -\frac{8\pi d_{1}}{(\varepsilon_{\rm He}+\varepsilon_{\rm d}) \bar{R}_{1||}^{2}}
      -\frac{8\pi d_{2}}{(\varepsilon_{\rm He}+\varepsilon_{\rm d}) \bar{R}_{2||}^{2}}
      -\frac{4\pi d_{3}}{\varepsilon_{\rm d} \bar{R}_{3||}^{2}},
    \label{6-4}     \end{equation}
      where $\varepsilon_{\rm He}$ and $\varepsilon_{\rm d}\simeq 3$ (see Ref.~\onlinecite{r3}) are the dielectric permittivities of helium and a
      dielectric, $d_{1}$, $d_{2},$ and $d_{3}$ are, respectively, the dipole moments of an
      atom in the first layer of helium and the first layer (near helium and a metal)
      of a dielectric, and $\bar{R}_{j||}$ are the
      mean longitudinal interatomic distances for these layers.
      For the first and second terms, we used the fact that a charge $q$ located in the first dielectric
       creates the potential
     $\varphi_{2} = \frac{2q}{(\varepsilon_{1}+\varepsilon_{2})R}$ in the second dielectric and
     $\varphi_{1} = \frac{q}{\varepsilon_{1}R} +
     \frac{(\varepsilon_{1}-\varepsilon_{2})q}{\varepsilon_{1}(\varepsilon_{1}+\varepsilon_{2})R^{\prime}}$
   in the first one (see  $\S 23$ in Ref.~\onlinecite{tamm}). In our case, $R^{\prime}=R$;
   therefore, $\varphi_{1} =
   \frac{2q}{(\varepsilon_{1}+\varepsilon_{2})R}=\varphi_{2}$.

Let us evaluate $a$. If $\nabla p=0$ in helium near the dielectric surface, then the helium-related part
$a$ is determined by formulas of Sec. IV B. But if $\nabla p \neq 0$, we take formulas of Sec. IV
A. The contribution to $a$  from the layers of dielectric atoms consists of the following parts: one proportional to $\alpha^{s}$ of the dielectric
(which must be small, since $\alpha^{s}$ of solids is usually much less than that of liquid helium) and one proportional to $\alpha$ of helium. The latter should be comparable with $a$ determined for helium in Sec. IV A or B.
This implies that $a$ should be of the order of magnitude of $a\approx -1$ from the first experiment \cite{rub}, \textit{but no exact coincidence is expected}.

Recently, we have read Ref. \cite{pash2010}, where a bulk model of polarization is proposed. Figure 2
of that work shows the experimental dependence of the signal in a second-sound wave on the coordinate $Z$ along a resonator. Though the
experimental data have not been published yet, we will make a preliminary comment. The points indicated in Ref.~\onlinecite{pash2010}
agree with formula (\ref{6-1}) describing the surface signal. The curve for the bulk signal differs from the experimental one by shape. The summary curve
(Fig.~\ref{fig4}) representing the surface and bulk signals does not differ by shape from the purely surface curve due to the smallness
of the bulk signal. Therefore, it is difficult to separate the bulk signal by the experimental points given in Ref.~\onlinecite{pash2010}.
However, the bulk contribution increases strongly, as the temperature increases and the ratio of the resonator length to its radius decreases.
With regard for this fact, it would be possible to determine the bulk signal, by using the full experimental data,
when they will be published. The data available at present indicate the mainly surface nature of the signal.

            \section{Conclusions}

    We have approximately calculated the electric
    signal $U$ arising at the electrode in the presence of a standing half-wave of first or second sound in He~II\@.
    The properties of the signal for second sound correspond to the
    experiment in its amplitude (approximately) and its independence on the resonator size and the temperature.
    Therefore, we have impression that the nature of the signal is clear on the whole.
     However, the properties of He~II near the metal surface are not clear in some aspects and are described only
     approximately. Therefore, the subsequent studies should be focused on this region.
      The signal for first sound has not been discovered experimentally till now, so that the formulas concerning
      first sound are the predictions of the model. In addition, the model predicts the formula of the dependence of the signal on the coordinate
      $z$ along a resonator and the strong growth of the signal for $^3$He--$^4$He mixture.
      It is shown also that the insignificant random admixtures ($\gsim 10^{-12}$) of polar molecules can completely
damp the bulk polarization of helium in first- and second-sound waves almost not affecting the surface polarization.
Therefore, we risk to assume that the bulk polarization is possible only in He~II specially purified from admixtures.

       In this connection, we indicate the problems of dry friction \cite{ginz1} and the exhaustion of
        $\rho_{s}$ at the wall. The consideration in Sec. III implies that the dry friction can
     appear below $T_{c} \sim 0.7\,$K. It should be of interest to measure it at all temperatures from $0$ to $T_{\lambda}$.

       To summarize, further studies are needed to attain complete understanding
       of the nature of the effect. In particular,
        is importantly to test experimentally the predictions of different models
        to determine the correct one.

               \section*{Acknowledgments}
     The author is  grateful to  A.\,B.~Kashuba, V.\,M.~Loktev and Yu.\,V.~Shtanov
     for useful discussions and to A.\,S.~Rybalko for his comments and some preliminary information about the new
     experiment. The author also thanks M.\,M.~Bogdan, E.\,Ya.~Rudavskii and the participants
     of the seminar at the Verkin Institute for Low Temperature Physics and Engineering
     for critical discussion of the results.

  \section*{Appendix A. Deja vu: the electric field in a spontaneously polarized dielectric}
We now discuss the methods of calculation of the electric field in
a ``spontaneously'' polarized dielectric such as He II with first-
or second-sound waves. In the literature, two different methods
leading to completely different results are developed. Therefore,
we will consider this question in more details.

In a number of works, in particular in one of the first works \cite{mel} and in new works \cite{pash2010,min2010}, the bulk
models of the electric activity of He II are considered without regard for the surface polarization. In Refs.~\onlinecite{mel,pash2010},
the polarization of helium is explained by that the atoms in a second-sound wave are accelerated and are polarized due to the inertia. The coupling of the
polarization and the acceleration $\textbf{w}$ is described by the Melnikovsky formula
 \begin{equation}
  \textbf{P} = -\gamma_{i}\textbf{w}.
        \label{7-1}     \end{equation}
According to Ref.~\onlinecite{min2010}, the polarization is related to collisions of atoms; as a result, formula (\ref{7-1})
acquires the different coefficient.
In these works, the potential is calculated by the formula
 \begin{equation}
  \textbf{E}=-\nabla\varphi,
        \label{7-2}     \end{equation}
and the field strength $\textbf{E}$ is determined by two means.  In Ref.~\onlinecite{pash2010}, it is described by the formula
 \begin{equation}
  \textbf{D} = \varepsilon \textbf{E}, \quad \textbf{P} = \kappa\textbf{E},
        \label{7-3}     \end{equation}
where $\kappa=(\varepsilon -1)/4\pi$ is the polarizability. In Refs.~\onlinecite{mel} and \onlinecite{min2010}, the field strength is determined from the condition
 \begin{equation}
  \textbf{D} = 0, \quad \textbf{E}=-4\pi\textbf{P}.
        \label{7-4}     \end{equation}
The proportionality $\textbf{E}= const\times \textbf{P}$ allows one to obtain a bulk electric signal that is of the order of the experimental
one and does not depend on the resonator size.
However, we do not agree with some positions of these works, in particular with the formula
$\textbf{w}=\partial (\textbf{v}_{n}-\textbf{v}_{s})/\partial t$ for the acceleration \cite{pash2010,min2010} and with the method of estimation of the coefficient $\gamma_{i}$
in formula (\ref{7-1}). The relation $\textbf{w}=\partial (\textbf{v}_{n}-\textbf{v}_{s})/\partial t$ is an
assumption, it allows one to obtain a signal weakly
depending on the temperature. By its physical sense, the quantity $\textbf{w}$ in formula (\ref{7-1}) is the local acceleration of a medium.
It is described by the formula $\textbf{w}=\frac{D}{Dt}\frac{\rho_{n}\textbf{v}_{n}+\rho_{s}\textbf{v}_{s}}{\rho}$
that is valid for first and second sounds.
The coefficient $\gamma_{i}$ was calculated in Refs.~\onlinecite{pash2010,min2010} in the first order of the stationary perturbation theory.
But, even in the calculation of the mutual (tidal) polarization of two immovable atoms, the \textit{second} order of perturbation theory
is used \cite{wb1,wb2,lt2}. The polarization arises at the acceleration due to the interaction of atoms. Therefore, it is one of the
manifestations of the tidal polarization of atoms. Respectively, it can be calculated for two atoms analogously to
Refs.~\onlinecite{wb1,wb2,lt2} in the second order of perturbation theory, but \textit{nonstationary}.  Hence we can estimate also the value
of $\gamma_{i}$ for the medium. Such a procedure of calculation of $\gamma_{i}$ seems more exact.

We now consider the methods of calculation of the electric field for the given problem in more details, since this question is of principal meaning.
We have He II with a
second-sound wave in the case where the polarization of a medium
(dielectric) is related to internal processes in the fluid, and no
external electromagnetic fields are present. Such a statement of
the problem is common for all authors. The key point is that without external field $\textbf{E}$ the source of the field
$\textbf{E}$ is the polarization of atoms. The problem can be
solved within two exact approaches: 1) to solve the Maxwell
equations
\begin{equation}
  {\rm div}\textbf{D}= 0, \quad {\rm rot}\textbf{E}=0
        \label{7-m}     \end{equation}
with regard for boundary conditions or 2) to calculate the potential as a sum $\varphi(\textbf{R}) = \sum\limits_{j}\frac{e_{j}}{|\textbf{R}-\textbf{r}_{j}|}$
of the potentials from all charges of the system (electrons and nuclei of the helium atoms). The helium atoms are not charged on the whole. Therefore, the potential is reduced to
the sum of the potentials from dipoles and higher multipoles of atoms. Since the contribution of dipoles is usually dominant, it is sufficiently
to sum over them:
   \begin{equation}
    \varphi (\textbf{R}) = n\int d\textbf{r}\frac{\textbf{d}(\textbf{R}-\textbf{r})}{\varepsilon
   |\textbf{R}-\textbf{r}|^{3}}.
               \label{7-5}  \end{equation}
Here, $\textbf{r}$ are the coordinates of atoms of helium, and $\textbf{R}$ is the observation point. Respectively, the field strength
  \begin{equation}
\textbf{E}(\textbf{R}) =-\nabla_{\textbf{R}}\varphi = \int d\textbf{r}\frac{3(\textbf{P}\textbf{n})\textbf{n}-\textbf{P}}{|\textbf{R}-\textbf{r}|^{3}},
                   \label{7-6}  \end{equation}
where $\textbf{P}=n\textbf{d}/\varepsilon $ is the polarization and $\textbf{n}=(\textbf{R}-\textbf{r})/|\textbf{R}-\textbf{r}|$.
The full polarization consists of spontaneous and induced parts:
 \begin{equation}
\textbf{P}=\textbf{P}_{sp}+\textbf{P}_{ind}, \quad \textbf{P}_{ind}=\kappa\textbf{E}.
                   \label{7-6b}  \end{equation}
In view of the smallness of $\kappa$ for He II,   the induced part is very small  and can be neglected, if the spontaneous polarization is the primary source of the field.
Such a method was used in Refs.~\onlinecite{lt1,gut,shev1,volpol} and above.

In Refs.~\onlinecite{mel,pash2010,min2010}, the calculation is performed in a different way with the use of relations (\ref{7-3}) or (\ref{7-4}). The results turn out
quite different. In particular, it was shown \cite{lt1,volpol} that the bulk signal depends strongly on the resonator
size. But no such dependence was found in Refs.~\onlinecite{mel,pash2010,min2010} and in experiments. Therefore, to clarify the nature of the signal,
it is important to know which method is more exact.

First, let us consider the applicability of the condition
$\textbf{D} = 0, \textbf{E}=-4\pi\textbf{P}$ in the bulk of
helium. As is known, this condition is valid on the boundary of a
dielectric and a metal. In the bulk, it is true only in several
cases. If a dielectric is surrounded by a resonator positioned in
an external field $\textbf{E}$, then the condition $\textbf{D} =
0$ is satisfied in the bulk in the absence of the spontaneous
polarization in a dielectric. In this case, we have also
$\textbf{E}=-4\pi\textbf{P}=0$. In our problem, the external field
is absent, and the polarization of helium is spontaneous.
Moreover, the condition $\textbf{D} = const$ can be satisfied
(which gives the relation $\textbf{E}=-4\pi\textbf{P}$ for the
variable field), if a dielectric is homogeneous in the direction
of the vector $\textbf{D}.$ In our case, the situation is
different, and the polarization is caused namely by the
inhomogeneity. This implies that the condition $\textbf{D} = 0$ is
not satisfied in the bulk for our problem. For the verification,
we use formula (\ref{7-6}) to determine $\textrm{div}\textbf{E}$
with regard for the relation
$\triangle_{\textbf{R}}|\textbf{R}-\textbf{r}|^{-1}=-4\pi\delta(\textbf{R}-\textbf{r})$.
We obtain $\textrm{div}\textbf{D}=0,$ i.e., the Maxwell equation
is satisfied. Analogously, we determine $\textbf{E}$, by assuming
for simplicity that the polarization $\textbf{P}$ is directed
along the $Z$ axis (as in the experiment). We obtain
 \begin{equation}
 \textbf{E}=-4\pi\textbf{P}+\tilde{E}\textbf{i}_{z}+\textbf{i}_{x}\int d\textbf{r}\frac{P(\textbf{r})\partial^{2}}{\partial R_{x}\partial R_{z}}\frac{1}
   {|\textbf{R}-\textbf{r}|} + (x\leftrightarrow y),
        \label{7-7}     \end{equation}
\begin{equation}
 \tilde{E}=\int d\textbf{r}P(\textbf{r})\left (\frac{\partial^{2}}{\partial R_{z}^{2}}-\triangle_{\textbf{R}}\right )\frac{1}
   {|\textbf{R}-\textbf{r}|}.
        \label{7-8}     \end{equation}
As is seen, the field is more complicated than that by the relation $\textbf{E}=-4\pi\textbf{P}.$ Indeed, we can separate the component
$-4\pi\textbf{P}$ from $\textbf{E}.$ But, in addition, we have nonzero components along the other axes and the component $\tilde{E}$ along the $Z$ axis.
The numerical analysis indicates (see Sec. VI) that last component is great. If we set $\textbf{E}= \nu\textbf{P}$, then the
coefficients of proportionality $\nu$ differ significantly from one another at different points and depend on the resonator size.
For example, at $R_{r}/L_{r}=0.2$ and $T=1.8$~K, we have $\nu \approx -2.5$ at the point in the middle of the resonator axis ($Z= L_{r}/2$),
and  $\nu \approx 1/\kappa \approx 220$ at the beginning of the axis ($Z\rightarrow 0$). Near the surface of the electrode, it is
necessary to consider the presence of several strongly polarized layers of helium. In this case, the field is perpendicular
to the surface, the condition $\textbf{D} = 0$ is satisfied, and it yields potential (\ref{5-7}), (\ref{new2}), (\ref{new4}).

In Ref.~\onlinecite{pash2010}, the field $\textbf{E}$ is determined from
formulas (\ref{7-3}). However, these formulas are valid only in
the case where a dielectric is polarized by an external field.
Indeed, the relation $\textbf{P} = \kappa\textbf{E}$ expresses the
fact that the electron shells of atoms of a dielectric are
stretched in an external field $\textbf{E}$, i.e., this relation
describes a response of the medium to an external field. In our
problem, no external field is present, and the source of the field
$\textbf{E}$ is the polarization of helium atoms. This polarization is related
to their interaction. We consider that the main contribution to
the polarization is given by the tidal polarization of atoms
arising in immovable atoms due to the interaction
\cite{wb1,wb2,lt2}. In Refs.~\onlinecite{mel,pash2010,min2010}, it is
considered that the polarization is related to a motion. It has no
meaning in the sense that the polarization is caused, in any case, by
processes in a fluid and is not connected with an external field.
Such a polarization is named  spontaneous. It is also observed
in piezoelectrics and pyroelectrics. In helium, a different
mechanism acts. At the spontaneous polarization, the relation
$\textbf{P} = \kappa\textbf{E}$ does not hold, obviously.

For clearness, we consider a simple example from the course of electrodynamics \cite{tamm,land8}: let us determine the field inside of a uniformly polarized
dielectric ball with radius $R_{0}$. We consider that the polarization $\textbf{P}$ is the same by direction and by magnitude
at all points of the ball.
The simple reasoning (see Ref.~\onlinecite{tamm}, $\S 24$, and Ref.~\onlinecite{land8}, $\S 13$, Exercise 1) allows one to write the answer
for the field at a point $\textbf{R}$:
   \begin{equation}
    \varphi (\textbf{R}) = 4\pi \textbf{P}\textbf{R}/3, \quad \textbf{E}(\textbf{R})=-4\pi \textbf{P}/3.
               \label{7-9}  \end{equation}
We can verify that the same result follows from the direct summation of the contribution of dipoles according to (\ref{7-5}) and (\ref{7-6}).
As is seen, the solution is obtained correctly in approach (\ref{7-5}), but relations (\ref{7-3}) and (\ref{7-4}) are not satisfied in such a system.
We note that, despite the proportionality of $\textbf{E}$ and $\textbf{P}$ in (\ref{7-9}), the connection between them is nonlocal (as distinct from
the local relations (\ref{7-3}) and (\ref{7-4})), since the coefficient of proportionality is determined by the contribution of all dipoles
of a dielectric from the region $r\leq R$.
Qualitatively, the problem for helium in a resonator is \textit{analogous}. But the polarization
in helium is inhomogeneous, a vessel is not sphere-like, and helium is bounded by a resonator; therefore, the connection between $\textbf{E}$ and $\textbf{P}$ is more complicated,
than (\ref{7-9}). To estimate the role of the inhomogeneity, we consider that the polarization of a ball increases proportionally to
$r$: $\textbf{P}(\textbf{r})=4r\textbf{P}_{0}/3R_{0}$ (in this case, the average over the bulk $\langle\textbf{P}\rangle=\textbf{P}_{0}$). Then the calculation gives
   \begin{equation}
    \varphi (\textbf{R}) =\pi \textbf{P}\textbf{R}, \quad \textbf{E}(\textbf{R})=-\pi (\textbf{P}+\textbf{n}(\textbf{n}\textbf{P})),
               \label{7-10}  \end{equation}
here $\textbf{P}\equiv\textbf{P}(\textbf{R})$ and $\textbf{n}=\textbf{R}/R$. In other words, the inhomogeneity leads to a change of the coefficient and to a complication of the dependence $\textbf{E}$ on
$\textbf{P}$. If the ball is stretched into a ``cigar'', then the dependence of the field on the ratio of the sizes of a cigar appears. In the ball, the field
from external uniformly polarized layers is equal to zero, since the layers are spherical. But they are nonspherical in the cigar, therefore the field
is nonzero, and the dependence on sizes of the system appears. If a dielectric is covered by a metal,
the field changes due to the contribution of images in a metal.

We note one more point. In approaches (\ref{7-2})--(\ref{7-4}),
the potential difference is determined as the integral
${\sss \triangle}\varphi = -\int\textbf{E}\textbf{ds}$. In Refs.~\onlinecite{mel,pash2010,min2010}, the contour is chosen along the
resonator axis $z$ between points on the resonator ends. In the
experiment \cite{rub}, one electrode is positioned on the
resonator end, and the second electrode is the metallic resonator
itself. In this case, the potential difference between the end
electrode and any point of the resonator is the same. But if the
potential difference is calculated by the relations
${\sss \triangle}\varphi = -\int\textbf{E}\textbf{ds}$ and (\ref{7-3}) or (\ref{7-4}), by positioning
the contour beginning on the first electrode and the contour end on
the internal surface of the resonator, then ${\sss \triangle}\varphi$ depends on the
coordinate $z$ of a point of the resonator. In particular, for the
points on the lateral surface with the coordinates $Z=0, 0.5L_{r},$
and $L_{r},$ the potential difference is equal to 0, 1, and 2 (in
arbitrary units).
This circumstance indicates that the field
$\textbf{E}$ is determined inaccurately. The equivalence of
different contours requires, in particular, that the field have
the great $\rho$-component. The approach in (\ref{7-5}) has the
same difficulty: it follows from formulas (\ref{4-7})--(\ref{4-9})
and Fig.~\ref{fig3} that the bulk potential difference between
the end electrode and points on the lateral surface of a resonator
with $Z=0, 0.5L_{r},$ and $L_{r}$ is equal to 0.16, 0.49, and
0.82. Such a distinction is related to the neglect of the images
of charges in a metal. The images must significantly affect the
bulk potential, but the difficulty consists in the determination of
a solution for bulk images. However, the main contribution to the
signal in our approach is a surface one, and it was determined
with regard for the images. Therefore, the neglect of images for
the bulk signal affects slightly the result. On the whole, the
structure of the field $\textbf{E}$ in the approach in (\ref{7-6})
is calculated much more exactly. Moreover, the dependence of the
signal on the resonator sizes is sensed and discovered. The consideration of
images should not exclude, obviously, this dependence.

This reasoning implies that the correct calculation of the field in a resonator is a sufficiently complicated problem. In our opinion,
formulas (\ref{7-5})--(\ref{7-6b}) lead to a significantly more exact result, qualitatively and quantitatively, than linear relations of
the form $\textbf{E}= const\times \textbf{P}$.

  \section*{Appendix B. Polarizability of liquid $^4$He}

         Of interest is  the question about how the presence of a tidal DM of He II atoms is manifested in properties of $\varepsilon (T,\rho)$.
          The Clausius--Mossotti relation
     \begin{equation}
   \varepsilon - 1 = \frac{4\pi\rho}{3}\frac{A}{M}(\varepsilon +2)
    \label{b-1}     \end{equation}
      implies that $\varepsilon$ depends on $T$ via the density $\rho
     (T)$ and the polarizability $A(T)$. The dependence $A(T)$ was measured in Refs.~\onlinecite{pol1,pol2}, where it was determined that
     $A$ decreases smoothly with increase in $T$ in the interval
     $T\approx 1.5\div 2.7\,$K, has a break at  $T=T_{\lambda}$, and increases with
     $T$ in the interval $T\approx 2.7\div 3.8\,$K. In this case, at densities
     corresponding to $T\approx 1.5\,{\rm K} \div T_{\lambda}$  and $T\approx 2.7\div 3.8\,$K, $A$ decreases with increase in
     $\rho$. This fact is qualitatively explained in Refs.~\onlinecite{a1,a2} in the frameworks of
     two mechanisms which, however, do not take the mutual
     polarization of atoms into account.
         We note that $A$ increases with $\rho$ at $T\approx T_{\lambda}\div
     2.7\,$K, which remains unclear.

      It was indicated in Ref.~\onlinecite{sht} that, at He II temperatures, $A$ depends on $T$
        approximately as
       \begin{equation}
  A \approx A_{0}(1+\delta_{0}/T).
    \label{b-2}     \end{equation}
    This relation corresponds to the Langevin--Debye law \cite{tamm}
      \begin{equation}
   \frac{\varepsilon - 1}{\varepsilon +2} = \frac{4\pi
   n d_{\rm in}^{2}}{9k_{B}T} + A_{0}\frac{4\pi\rho}{3M}
   \label{b-3}     \end{equation}
    that is valid for a gas of molecules possessing the intrinsic DM $d_{\rm in}$. Therefore, it was assumed in Ref.~\onlinecite{sht} that
     the dependence $A(T)$ (\ref{b-2}) is related to the intrinsic DM
     of helium atoms. This gives the alternative explanation of the dependence $A(T)$
     for $T \leq T_{\lambda}$.

      The data $A(T=1.5\,{\rm K}) = 0.1232\,{\rm cm}^{3}/\mbox{mole}$ and
      $A(T=2.0\,{\rm K}) = 0.12305\,\mbox{cm}^{3}/\mbox{mole}$ (see   Ref.~\onlinecite{pol2}) yield
      $ A_{0}\approx  0.1226\,\mbox{cm}^{3}/\mbox{mole}$ and $\delta_{0} \approx
      9{\rm K}/1226$. Since $\varepsilon = 1.057195$ (see Ref.~\onlinecite{pol1})  and
       $\rho = 0.14526\,\mbox{g}/\mbox{cm}^{3}$ at $T=1.63\,$K (see Ref.~\onlinecite{pol2}), relation (\ref{b-1}) yields
     \begin{equation}
  \varepsilon = \varepsilon_{0}\left (1+\frac{\beta_{0}\delta_{0}}{T}\right )
    + \frac{4\pi A}{3M}(\varepsilon +2)(\rho (T)-\rho(1.63\,{\rm K})),
     \label{b-5}     \end{equation}
      where $\varepsilon_{0}\approx 1.057$ and $\beta_{0} \approx 0.0539$.
          According to the modern
     theory, a free $^4$He atom has no intrinsic DM. However, in the environment of other
     helium atoms, the atom acquires a fluctuating tidal DM with the mean modulus
     $\tilde{d}$. The shape of the function $g(r)$  (see Ref.~\onlinecite{ss}) for He II
     testifies that the mean deviation of the interatomic
     distance in He II from $\bar{R}$ is about $\delta R \simeq \bar{R}/6$. Let the distance from the
     given He II atom to the left adjacent atom be $\bar{R}-\delta R$, and let the distance to the
     right one be $\bar{R}$. We also mention another characteristic configuration with
     the distances $\bar{R}+\delta R$ and $\bar{R}$. According to
     (\ref{3-6}) and (\ref{3-7}), these neighbors induce the DM $d_{x}^1 = d_0 \left (
     \frac{\bar{R}^{7}}{(\bar{R}-\delta R)^{7}} - 1 \right )$ on the atom for the
     first configuration and
     $d_{x}^2 = d_0 \left ( 1-\frac{\bar{R}^7}{(\bar{R}+\delta R)^{7}} \right
     )$ for the second one, and the mean value is $d_{x}\simeq
     (d_{x}^{1}+d_{x}^{2})/2\approx 1.6d_{0}$.  In the
  $Y$ and $Z$ directions, the induced DM of the atom is the same, so that the mean modulus
    of the total DM of the atom is $\tilde{d}\approx\sqrt{3}d_{x}\approx 2.8d_{0}$. Such an estimate is analogous to the derivation of formula (17) in Ref.~\onlinecite{volpol}. The consideration of the structural factor $S(k)$ leads to formula (30) in Ref.~\onlinecite{volpol}, where the coefficient is greater by a factor of 3.24.
     So, we should increase $\tilde{d}$ by approximately the same factor in order to take $S(k)$ into account. Finally, we get
    \begin{equation}
  \tilde{d} \approx 2.8\cdot 3.24d_{0}  \approx 3.22\cdot 10^{-4} |e| a_{B}
   \approx 2.73\cdot 10^{-33}\mbox{C}\cdot\mbox{m}.
  \label{b-6}     \end{equation}
    The quantity $\tilde{d}$ turns out to be greater than $|d_{\rm h}|$ (\ref{3-11}):  $\tilde{d}\approx
    -2.44 d_{\rm h}$.  At the same time, relations (\ref{b-1})--(\ref{b-3})
   yield $d_{\rm in}\approx 3.2\cdot
   10^{-33}\mbox{C}\cdot\mbox{m}\approx 1.17\tilde{d}$, i.e. $d_{\rm in}$ is close to $\tilde{d}$.

    We note that the DM vector of an atom averaged over the time is zero. But the mean DM modulus is nonzero,
     and the Langevin--Debye formula (\ref{b-3}) is valid just for the nonzero DM modulus \cite{deb}.

    For fluids, we should add the factor
     $q=(1-i\omega\tau)^{-1}$ (see Ref.~\onlinecite{deb}) to the right-hand side of (\ref{b-3}), where $\omega$ is the external field frequency,
     and $\tau = 8\pi\eta \tilde{a}^{3}/(k_{B}T)$ is the relaxation time. For He II, we obtain $\tau\sim 10^{-13}\mbox{sec}^{-1}$ and $\omega\tau \sim 10$
     (for experimental microwaves \cite{pol1}). But this is true for polar molecules, whereas
     the DM of helium atoms is not intrinsic, but tidal. In the latter case, the value of $\tau$
     is unknown, but it is probably much less due to the tough coupling with
     adjacent molecules, so that  $q\approx 1$. We may conclude that
     formula (\ref{b-3}) is valid at $\omega\tau \ll 1$ for liquid helium as well.
          Since $d_{\rm in}\approx \tilde{d}$, it is possible that $q\approx 1$ for experimental values of $\omega$. Then
      the dependence  $\varepsilon(T)$ at $T \leq T_{\lambda}$ can be explained by the tidal DM of He II atoms.

      Because $\tilde{d} \sim \delta R/\bar{R}^{8}\sim \bar{R}^{-7}\sim \rho^{7/3}$ and $\delta_{0} \sim \tilde{d}^2$,
      we obtain that relation (\ref{b-2}) can be written more exactly as
          \begin{equation}
  A \approx   A_{0}\left (1+\frac{\delta_{0}\cdot\rho^{14/3}(T)}{T\cdot\rho^{14/3}(1.63\,{\rm K})}\right ).
    \label{b-7}     \end{equation}

      Let us turn to the experimental dependence $A(T)$.
      It was assumed in Refs.~\onlinecite{pol1,pol2} that, at the temperatures $T\approx 1.5\,{\rm K} \div T_{\lambda}$  and $T\approx 2.7\div 3.8\,$K,
      where $A$ decreases with increase in
     $\rho$, the dependencies $A(T)$ and $A(\rho)$ are determined \cite{a1,a2}
     by a decrease in the distance between atoms with increase in $\rho$.
     In this case, the increase in $A$ with $\rho$ at  $T\approx T_{\lambda}\div
     2.7\,$K remains to be unclear. According to Ref.~\onlinecite{sht}, the dependencies $A(T)$ and  $A(\rho)$ on the
     interval  $T\approx 1.5\,{\rm K} \div T_{\lambda}$ can be explained
     in a different way with the use of formulas (\ref{b-2}) and (\ref{b-3}). We consider that,
      for all $T$ and $\rho$, both mechanisms (a change in the density and the Langevin--Debye mechanism)
      contribute to the dependencies $A(T)$ and  $A(\rho)$. In this case, the Langevin--Debye mechanism is related to the mutual polarization of atoms.
      If $q\approx 1$, then the main contribution to $A(T)$ at $T < 2.7\,$K is given by the
      Langevin--Debye mechanism by (\ref{b-2}) and (\ref{b-7}). As $T$ increases, this contribution decreases and, at $T > 2.7\,$K,
      becomes less than the addition due to a change in the density. The latter
      is not taken into account in (\ref{b-2}) and (\ref{b-7}), but it determines the dependences $A(T)$ and $A(\rho)$ at $T > 2.7\,$K.
      It is easy to verify that formulas (\ref{b-2}) and (\ref{b-7}) well describe the experimental behavior of $A(T)$ and $A(\rho)$
      (see Refs.~\onlinecite{pol1,pol2}) at all $T < 2.7\,$K,  except for the break at $T= T_{\lambda}$. Thus, the behavior of
      $A(T)$ and  $A(\rho)$ at $T\approx T_{\lambda}\div
     2.7\,$K, where $A$ increases with $\rho$, is explained as well. We note that it is difficult to determine which of formulas
     (\ref{b-2}) or (\ref{b-7}) corresponds better to the
     experiment \cite{pol1,pol2}.
       If, nevertheless, $\omega\tau \gg 1$, then $q \ll 1$, and
       the Langevin--Debye mechanism leads to $A(T)$ much less than the experimental value and fails to describe the experimental $A(T)$ at $T < 2.7\,$K.

        \end{document}